\documentclass[fleqn,usenatbib]{mnras}

\usepackage{newtxtext,newtxmath}

\usepackage[T1]{fontenc}

\DeclareRobustCommand{\VAN}[3]{#2}
\let\VANthebibliography\thebibliography
\def\thebibliography{\DeclareRobustCommand{\VAN}[3]{##3}\VANthebibliography}

\usepackage{graphicx}	
\usepackage{amsmath}	

\title[$G^3C$v$_{11}$ and updated HMF]{Galaxy And Mass Assembly (GAMA): Updated GAMA Group Catalogue ($G^{3}C$v$_{11}$) and Halo Mass Function using improved, uniform photometry from KiDS} 

\author[T. S. Lambert et al.]{
  Trystan S. Lambert,$^{1}$\thanks{E-mail: trystan.lambert@uwa.edu.au}
  S. P. Driver,$^{1}$
  A. S. G. Robotham $^{1}$
  and S. Bellstedt $^{1}$
  \\
  $^{1}$ International Centre for Radio Astronomy Research (ICRAR), University of Western Australia, Crawley, WA 6009, Australia\\
}

\date{Accepted XXX. Received YYY; in original form ZZZ}

\pubyear{\the\year{}}

\begin{document}
\label{firstpage}
\pagerange{\pageref{firstpage}--\pageref{lastpage}}
\maketitle

\begin{abstract}
We present an updated Galaxy and Mass Assembly (GAMA) Galaxy Group Catalogue---$G^{3}C$v$_{11}$ which is built from the most recent GAMA data release, using updated and consistent photometry from KiDS across all primary fields. $G^{3}C$v$_{11}$ also includes the G23 field which has not been publicly available in previous versions of the GAMA group catalogue. In addition, we also present a new group catalogue for the Sloan Digital Sky Survey using ``Nessie'', the same group finding tool we used for GAMA. We use these two consistently defined group catalogues, along with Halo Mass Function (HMF) measurements from REFLEX II, to build an empirically measured HMF and find that the general shape of the HMF closely follows what is predicted from $\Lambda$CDM.
\end{abstract}




\section{Introduction}
Galaxies are not uniformly distributed within the universe; they are instead found to organize into large-scale structures such as filaments, groups, voids, and clusters \citep{bond1996, jarrett2004, libeskind2018}. This is clearly seen when looking at the distribution of galaxies and has been evident since the very first systematic redshift surveys, such as the Center for Astrophysics (CfA) \citep{Huchra1983} with only 2401 galaxies. This has only become clearer through the decades through various efforts to map out the 3D distribution of galaxies. Notable examples include the \textit{2-degree-Field Galaxy Redshift Survey} \citep[2dFGRS;][]{colless2001}, \textit{6-degree-Field Galaxy Survey} \citep[6dFGS;][]{jones2004}, \textit{Sloan Digital Sky Survey} \citep[SDSS;][]{york2000}, Galaxy and Mass Assembly \citep[GAMA;][]{driver2011}, the \textit{2MASS Redshift Survey} \citep[2MRS;][]{huchra2012, macri2019}, and the Dark Energy Spectroscopic Instrument \citep[DESI;][]{DESIdr1}. 

Besides being visually obvious, these large-scale structures are frequently identified within these redshift surveys using a variety of algorithmic techniques and multiple catalogues have been produced for filaments \citep[e.g.][]{alpaslan2014, Tempel2014Filaments} and voids \citep[e.g.][]{Sutter2012}. However, the most common, and arguably most important, large-scale structure catalogues produced from redshift surveys are the groups and clusters \citep[e.g.][]{huchra1982, yang2005, robotham2011, tempel2014, tempel2017, lambert2020}. These are usually the first large-scale structure catalogues to be produced and have proven to be the dominant environmental variable when discussing galaxy evolution \citep{McGee2009}.

The existence of these structures is expected and is a natural consequence of the hierarchical model of structure formation from $\Lambda$CDM cosmology \citep{peebles1980}. Indeed, $\Lambda$CDM makes predictions of the statistical distribution of these structures \citep{Murray2021}, and therefore their observations provide a test for $\Lambda$CDM. One particular observable which can be used to test $\Lambda$CDM is by comparing the distribution of Halo Masses, i.e., the Halo Mass Function (HMF) to that predicted by $\Lambda$CDM \citep{Murray2013, Murray2021}. 

\cite{driver2022a} did this exact test using the mass estimates from groups in the GAMA group catalogue \citep{robotham2011}, SDSS group catalogue \citep{tempel2017} and REFLEX II \citep{Boringer2013, Boringer2017} to empirically construct the HMF and compare it to the predictions made by \cite{Murray2013}.

The latest GAMA release \citep[DR4;][]{driver2022}, included three substantial photometric changes from previous data releases (see \citealt{bellstedt2020} for explicit details): First, DR4 switched the photometry from SDSS to the Kilo Degree Survey \citep[KiDS;][]{dejong2013} photometry which was consistent across all fields; second, GAMA sources were reidentified using ProFound \citep{Robotham2018} replacing the previous sources which were identified using SExtractor \citep{Bertin1996}; third, while previous GAMA data releases had more than 98$\%$ completeness at $r<19.8$, the new photometry lowered this to $r<19.65$.

Even though DR4 was available, no group catalogue was built for this release, and the most recent public release of the group catalogue ($G^{3}C$v$_{10}$) did not include the G23 field. Therefore, in constructing the HMF, \cite{driver2022a} only used the three complete equatorial fields (G09, G12, and G15) using the old SDSS photometry, effectively using DR3 without G23 \citep{baldry2018}. 

In addition to these changes, a generalized group finding tool---Nessie \citep{lambert2026}---was recently released and is built upon the original friends-of-friends group finder used in \cite{robotham2011}. With the advent of Nessie, we are well placed to 1) publish the final GAMA group catalogue ($G^{3}C$v$_{11}$) using the most recent photometry and latest GAMA data release from \cite{driver2022} and 2) update the HMF published in  \cite{driver2022a} using these new catalogues. 

The remainder of this paper is organized as follows: in Section 2 we describe the GAMA and SDSS galaxy catalogues that we use to build our groups as well as the REFLEX II data points that we incorporate into the HMF; in Section 3 we give a brief explanation of the friends-of-friends algorithm as used by Nessie to build the group catalogues, the mass estimators used in each catalogue, as well as the techniques used in \cite{driver2022a} to construct the HMF from these masses; In Section 4 we present two new group catalogues---$G^{3}C$v$_{11}$ and the Nessie SDSS catalogue---as well as the resulting HMF fits; we discuss these results in Section 5 and conclude in Section 6. Throughout this paper we make a concerted effort to convert all cosmology values to a standard Planck cosmology \citep{plank2018} assuming $H_0 = 67$ km s$^{-1}$ Mpc$^{-1}$ and $\Omega_M=0.315$. We state any differences in cosmology explicitly. All magnitudes are in the AB system.

\section{Data}
\begin{figure*}
  \includegraphics[width=\textwidth]{"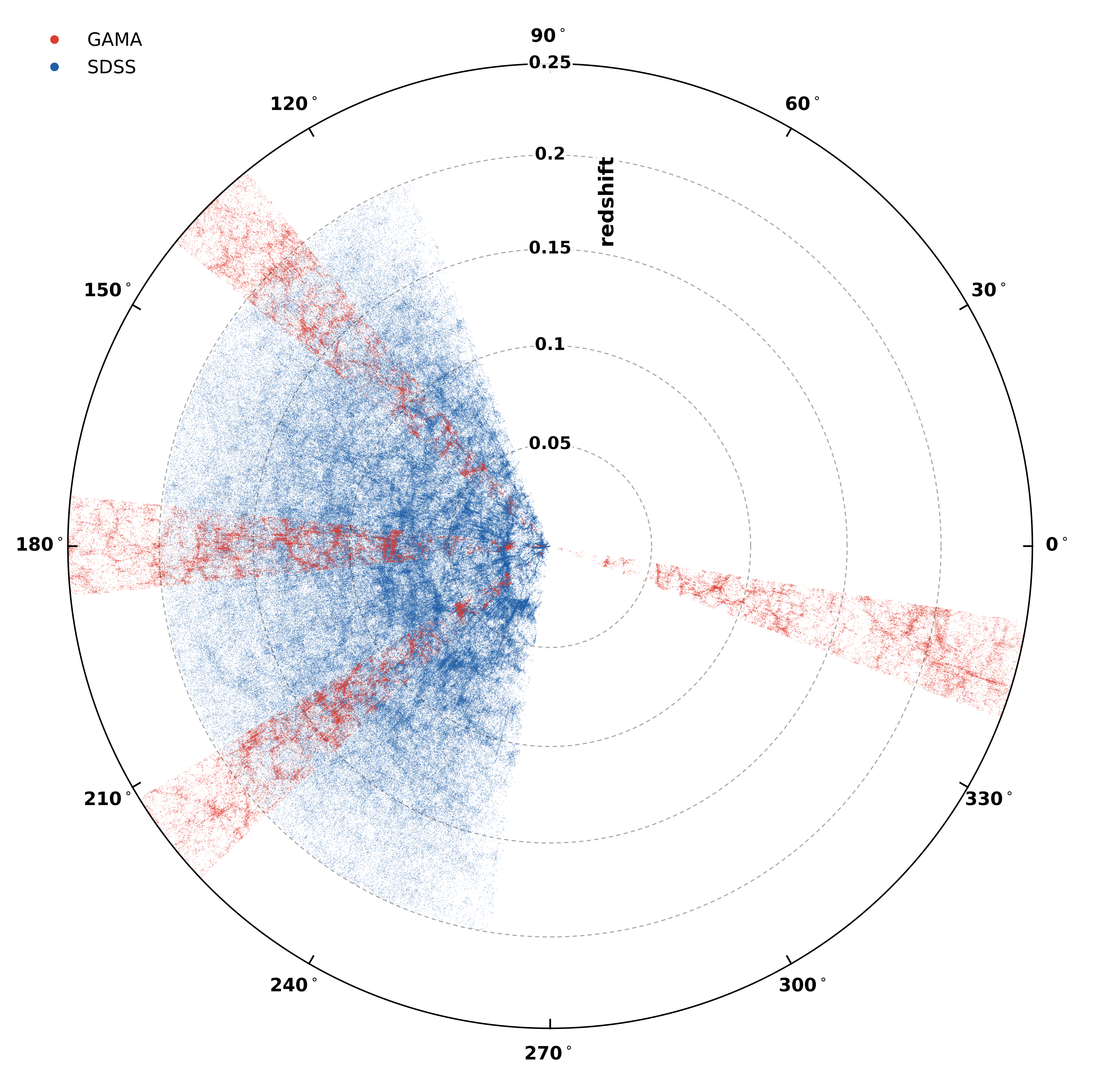"}
  \caption{Polar scatter plot showing the large-scale distribution of galaxies in the SDSS (blue) and GAMA survey (red). RA is shown in the angular direction whilst redshift is represented in the radial direction, with the dashed rings marking intervals of $\Delta z = 0.05$ (labelled along the vertical axis).}  
  \label{fig: on-sky plot}
\end{figure*}

This work relies on three main data sets: the two redshift surveys---GAMA \citep{driver2022} and SDSS main \citep{alam2015}---from which we build group catalogues to use in the HMF; and the HMF points from the ROSAT-ESO Flux Limited X-ray Galaxy Cluster Survey \citep[REFLEX II;][]{Boringer2017}.

Both the redshift surveys are presented in Fig. \ref{fig: on-sky plot} as a polar plot, with RA in the angular dimension, and redshift in the radial. The red points show the GAMA galaxies whilst blue shows SDSS. Even though we build the group catalogues for larger redshifts, we only show galaxies up to $z<0.25$ since these were the ones that we restrict ourselves to when building the HMF. Both survey designs are clear, with SDSS having a much larger area but being shallower, whereas GAMA is focused on individual, deeper fields, which span across the sky to combat cosmic variance. 

\subsection{GAMA DR4}
The GAMA survey is a spectroscopic redshift survey which aimed to observe five $\sim$60 deg$^2$ regions to a completeness limit of $r<19.8$ with 98$\%$ completeness \citep{driver2022}. These being the five GAMA regions, namely:  G02 ($30.2^{\circ} < \alpha < 38.8^{\circ}$; $-10.25^{\circ} < \delta<  -3.72^{\circ}$), G09 ($129^{\circ}< \alpha < 141^{\circ}$; $-2^{\circ} < \delta < 3^{\circ}$), G12 ($174^{\circ} < \alpha < 186^{\circ}$; $-3^{\circ} < \delta < 2^{\circ}$), G15 ($211.5^{\circ} < \alpha < 223.5^{\circ}$; $-2^{\circ} < \delta < 3^{\circ}$), and G23 ($339.0^{\circ} < \alpha < 351.0^{\circ}$; $-35^{\circ} < \delta < -30^{\circ}$). 

Four of the five fields (G09, G12, G15, G23) fall within the European Southern Observatory (ESO) VST KiDS and VISTA VIKING footprints. As such, the latest data release of GAMA \citep[DR4; ][]{driver2022} included updated photometry for all the objects in these fields as well as improved source identification \citep{bellstedt2020}. This was an improvement on previous data releases of GAMA since the inclusion of KiDS photometry allows for all four primary fields to have improved photometry and uniform imaging, as well as better source identification of the targets by making use of the ProFound software package \citep{Robotham2018} which, amongst other things, takes into account irregular shapes and watershed deblending \citep{bellstedt2020}. 

Since G02 has very different photometric coverage and was incomplete, it is often not included as part of the core GAMA survey. We adopt the same precedent and only use the four premier fields (G09, G12, G15, G23), with a completeness of >95$\%$. This results in a final catalogue of 195 432 galaxies. 

These data were downloaded from the GAMA survey DR4 schema browser\footnote{\url{https://www.gama-survey.org/dr4/schema/}}, from the science ready input catalogue which contains the new ProFound photometry (gkvScienceCatv02). From this table we select galaxies with a science-quality redshift (normalized redshift quality > 2), non-masked sources, and sources that are not marked as duplicates because of overlapping tiles\footnote{SELECT UberID,CATAID,RAcen,Deccen,flux$\_$rt, flux$\_$err$\_$rt, Z FROM gkvScienceCatv02 WHERE NQ>2 AND SC>6 AND duplicate=0 AND mask=0 AND starmask=0}.

\subsection{SDSS}
\cite{driver2022a} used the Sloan Digital Sky Survey (SDSS) to help constrain the HMF. This was done by using the group catalogue constructed by \cite{tempel2017}. We also include SDSS to help constrain the HMF, however, in order to make the results as consistent as possible, we choose to build our own group catalogue from the same galaxy sample that was used in \cite{tempel2017} instead of using their group catalogue. 

\cite{tempel2017} use SDSS Data Release 12 \citep{eisenstein2011,alam2015} (the blue continuous survey in Fig. \ref{fig: on-sky plot}) and make their galaxy sample readily available as a downloadable catalogue through VizieR\footnote{\url{https://vizier.unistra.fr/}}. This SDSS galaxy sample was built from galaxies within the SDSS legacy region, which were flagged as galaxies or quasars in the spectroscopic catalogue as well as galaxies in the photometric catalogue, with magnitudes brighter than $r<17.77$ and redshift less than 0.2. The $z<0.2$ limit is therefore part of the \cite{tempel2017} sample definition, which we inherit unchanged by using their galaxy table; it is not a choice made in this work. Some erroneous entries were visually inspected and removed and an additional 1349 galaxies were added from \cite{tempel2014}. The final number of galaxies was 584 449 (for explicit details see \citealt{tempel2014} and \citealt{tempel2017}).  

\subsection{REFLEX II}
In addition to the two group catalogues that we generate from GAMA and SDSS, we also include additional data points from the HMF determined in \cite{Boringer2017} which was built using the REFLEX II galaxy cluster survey \citep{Boringer2013}. This cluster catalogue was built from the ROSAT all-sky survey, covering a 4.24 steradian area in the region defined by $\delta < 2.5^{\circ}$ and excluding the galactic plane with a galactic latitude of $|b| \geq 20^{\circ}$, and consisted of 911 X-ray luminous clusters.

\section{Method}
\subsection{Building the group catalogue}

GAMA has benefited greatly from improved source identification and better photometry. However, these improvements come at the cost of differing input catalogue and magnitude limits which requires updating the group catalogue. Whilst we do not expect the changes in the photometry to significantly change the overall results, we do take this opportunity to build a new catalogue with the improved photometry and include the G23 field in its release.

\subsubsection{Group finder}
We build the new group catalogue by making use of the new friends-of-friends galaxy group finder, Nessie \citep{lambert2026}. Nessie was built from the original group finder used in \cite{robotham2011} and is therefore an optimal tool to use for building the GAMA group catalogue. It is worth noting that Nessie recovers the exact same results as the group finder used in \cite{robotham2011} after making two explicit parameter choices: Firstly, the linking lengths in \cite{robotham2011} rely on a tuning of $E_b$ and $E_R$ which influence the on-sky and radial linking lengths respectively, however, they found in that paper that these parameters had negligible effect and had optimum fitted values of 0. Therefore, Nessie makes this assumption and fixes both of these parameters to zero. Secondly, \cite{robotham2011} introduce a softening power, $\nu$, which effectively increases the linking lengths for intrinsically brighter galaxies. Whilst this addition makes intuitive sense, practically this parameter does not seem to have much of an effect. Indeed, $\nu$ was found to be the least impactful parameter of all of those introduced in \cite{robotham2011}. This is driven by the fact that the galaxies that would have had their linking lengths boosted the most (i.e., the brightest galaxies) would likely already find themselves in a group or at least with a nearby neighbour, meaning that the larger linking length may not actually result in a different group being identified. Therefore, Nessie also sets $\nu=0$. 

These two parameter choices result in a simpler definition of the linking lengths used to determine if two galaxies are ``friends''. The final linking length definition which determines if galaxy $i$ is friends with galaxy $j$ is defined as:

\begin{equation}
  \tan(\theta_{ij})(D_{c,i} + D_{c, j}) \leq D_i + D_j,
\end{equation}
and
\begin{equation}
  |D_{c, i} - D_{c,j}| \leq \frac{V_i + V_j}{2},
\end{equation}
where $\theta_{ij}$ is the angular separation between galaxy $i$ and $j$, and $D_i$ and $V_i$ are the on-sky and line-of-sight linking lengths for galaxy $i$, defined as:
\begin{align}
  D_i &= b_0\left[\frac{1}{\bar{\rho}(z_i)c_i}\right]^{1/3},\\
  V_i &= R_0D_i,
\end{align}
where $b_0$ and $R_0$ are free parameters to be fit, $\bar{\rho}(z_i)$ is the average local density at the redshift of galaxy $i$, and $c_i$ is the completeness around galaxy $i$, which we calculated to be the completeness in a sphere around galaxy $i$, with a comoving radius of 1Mpc.

Nessie comes with several helper functions to make reasonable choices for $b_0$ and $R_0$ as well as defining the $\bar{\rho}(z)$ function, including routines for tuning against mock catalogues, and building $\bar{\rho}(z)$ from the observed $n(z)$ distribution using randoms catalogues.

\subsubsection{Tuning linking lengths}
One of the core difficulties in using a FoF based algorithm is determining the choice for linking lengths. There are several approaches one could take. \cite{crook2007} determined the linking lengths analytically but then also used a much stricter pair of linking lengths and published both a low and high density group catalogue. \cite{lambert2020} iteratively built catalogues using stricter linking lengths and then selected the optimum set of linking lengths through visual identification in virtual reality \citep{jarrett2021}. \cite{tempel2016} calibrated their linking lengths using the mean distance between galaxies. Another very popular method is running an optimization routine and comparing results to a mock catalogue with known ``true'' groupings as done in \cite{robotham2011}. We adopt this latter method in keeping with the precedent set in previous editions of the GAMA group catalogue.

Nessie implements the same fitting routine used in \cite{robotham2011} to compare the groups found vs the known groupings from the mock catalogues. For completeness we include the formal description of the statistic and optimization, but refer the reader to both \cite{robotham2011} and \cite{lambert2026} for a more complete description.

A statistic, $S_{\rm total}$, is defined as the product of grouping efficiency ($E_{\rm total}$), and grouping purity ($Q_{\rm total}$),
\begin{equation}
  S_{\rm total} = E_{\rm total}Q_{\rm total},
\end{equation}
where
\begin{equation}
  E_{\rm total} = E_{\rm FoF}E_{\rm mock} = \left(\frac{N_{\rm bij}}{N_{\rm FoF}}\right)\left(\frac{N_{\rm bij}}{N_{\rm mock}}\right)
\end{equation}
and
\begin{equation}
  Q_{\rm total} = \left(\frac{\sum\limits_{i=1}^{N_{\rm FoF}} (P_{\rm FoF,i}) (n_{\rm FoF,i})}{G_{\rm FoF}}\right)\left(\frac{\sum\limits_{i=1}^{N_{\rm mock}} (P_{\rm mock,i}) (n_{\rm mock,i})}{G_{\rm mock}}\right)
\end{equation}
where $N_{\rm  bij}$ is the number of bijective groups across the mock and constructed group catalogues (defined to be groups that share $\geq 50\%$ of their members), $N_{\rm FoF}$ and $N_{\rm mock}$ are the total number of groups in the constructed and mock catalogues, $G_{\rm FoF}$ and $G_{\rm mock}$ are the total number of galaxies in groups in the constructed and mock catalogues, $P_{\rm FoF,i}$ and $P_{\rm mock, i}$ are the purity products of the i$^{\rm th}$ best-matching FoF and mock group, and $n_{\rm mock, i}$ and $n_{\rm FoF, i}$ are the number of members in the overlapping group $i$ in the mock catalogue and constructed catalogue respectively.

Since Nessie is derived from the GAMA group finder of \cite{robotham2011}, we retune the algorithm on the same mock catalogues used in that work in order to establish an appropriate set of linking parameters. Using the GAMA DR1 mock lightcones constructed from the Millennium simulation, we obtain best-fit values of $b_0 = 0.04 \pm 0.01$ and $R_0 = 36\pm4$ with a total score of $S_{\rm total} = 0.38$.

These values differ from those reported by \cite{robotham2011}, who found $b_0 = 0.06$ and $R_0=18$, but the resulting score is comparable to their $S_{\rm total} = 0.41$. Given that the plane-of-sky and radial linking lengths are coupled, and that the algorithm has undergone minor modifications, such differences are expected. The close agreement in $S_{\rm total}$ indicates that the adopted parameters are appropriate and that Nessie recovers group catalogues of comparable quality to the original implementation (see \cite{lambert2026}). Furthermore, we find that the actual group finding results are quite robust to the particular choice of linking lengths, in particular the radial linking length has a large range of ``equally good'' values due to the fingers-of-god effect; this parameter can be somewhat degenerate, so long as the radial linking length is shorter than the radial inter-group distance, and longer than the inter-galaxy distance, then the same group will be recovered.

\subsubsection{Mass estimates}
A key component for determining the HMF is how to estimate the mass of the groups. This can be done using X-ray observations \citep[e.g.][]{Boringer2017} or weak-lensing \citep[e.g.][]{viola2015} but, in redshift surveys, this is most often done by inferring the virial mass of the groups using their velocity dispersion \citep{eke2004, crook2007, robotham2011, tempel2014, tempel2016, lambert2020, vanKempen2025, lambert2026}. However, mass can be calculated using a variety of estimators, which rely on the overall velocity dispersion of the galaxies. All these properties can be estimated in a variety of ways which means that the final mass estimate of a given group catalogue is often dependent on specific choices made throughout the work. Moreover, the group mass accuracy is dominated by the specific membership identification within the group \citep{duarte2014}, i.e., no mass estimate is useful unless the group members have been correctly identified. Ultimately, the group mass in a group catalogue can become very ambiguous with even identically identified groups in different catalogues reporting different masses because of subtle choices made with regards to the calculation of the line-of-sight velocity dispersion.

Different surveys justify their particular choice of mass estimate either through comparisons to previous catalogues, externally validated mass estimates, comparison to simulations and mock catalogues, or a combination of all of these things \citep{eke2004, crook2007, robotham2011, tempel2014, lambert2020}. And while Nessie provides a variety of methods for determining mass and velocity dispersion, ultimately, we choose to adopt the same mass estimates used by the original authors of each respective group catalogue, i.e., we use the best attempted mass estimate from \cite{robotham2011} for GAMA, and the best attempted mass estimate from \cite{tempel2017} for SDSS. While the mass estimators are different, this is consistent with the approach taken to derive the HMF by \cite{driver2022a} where a variety of mass estimators were used. Furthermore, we get the additional benefit of using the mass estimates which these individual surveys have justified and been validated in subsequent works.

In GAMA the mass estimate comes from equation~18 of \cite{robotham2011}
\begin{equation}
  M = \frac{A}{G} \left(\sigma_{\rm FoF}\right)^2 r
\end{equation}
where $G$ is the gravitational constant, $\sigma_{\rm FoF}$ is the velocity dispersion, $r$ is the radius of the group, and $A$ is a scaling factor that is tuned to a mock catalogue. The choice of $A$ comes from comparisons to the GALFORM semi-analytic model \citep{bower2006}, and \cite{robotham2011} provide two mass estimates depending on the tuning of the $A$ parameter: either as an overall constant value, where they determine $A=10$ and is reported as \texttt{MassA} in $G^{3}C$v$_{10}$; or on a per-group basis by fitting a functional form to $A$ depending on the redshift and multiplicity of a group, provided as \texttt{MassAfunc}. This has the downside of tying these mass estimates to a particular mock catalogue, in this case GALFORM, and for this reason \cite{driver2022a} chose to instead adopt a constant $A$ value, but chosen from the mean ratio of both estimates, resulting in them adopting $A=13.9$. In this work we choose to adopt $A=10$ in keeping with the best results from \cite{robotham2011}. This makes $G^{3}C$v$_{11}$ consistent with previous versions of the catalogue. For the HMF, however, we instead follow \cite{driver2022a} and adopt $A=13.9$, with the multiplicity debiasing described in \S\ref{sec:mass-error}, so that our HMF is directly comparable to theirs. The masses used in the HMF are therefore not the \texttt{MassA} or \texttt{MassAfunc} columns of the released catalogue, but are recomputed from \texttt{VelDisp} and \texttt{Rad50}. 

\cite{robotham2011} follow \cite{eke2004} in determining the velocity dispersion of the group by using the ``gapper method'' \citep{beers1990} which is very popular and has become the de facto standard method of estimating the velocity dispersion of groups due to its robustness to low number statistics. 

\begin{equation}
  \sigma_{\rm gap} = \frac{\sqrt{\pi}}{N(N-1)}\sum\limits_{i=1}^{N-1}w_ig_i
\end{equation}
where $g_i$ is the difference in velocity (gap) between galaxy $i$ and $i+1$ where the galaxies have been sorted in ascending order according to velocity, $w_i = i(N-i)$, and $N$ is the number of members in that group.

Both \cite{eke2004} and \cite{robotham2011} include an additional correction which increases the velocity dispersion estimate to account for the coupling between the brightest galaxies and the group centre. The final value used in determining $\sigma_{\rm FoF}$ for a group is:
\begin{equation}
  \sigma = \sqrt{\frac{N}{N-1}\sigma_{\rm gap}^2 - \sigma_{\rm err}^2}
\end{equation}
where $\sigma_{\rm err}$ is the sum in quadrature of the velocity error for each member of the group.

The mass estimates in the \cite{tempel2017} SDSS catalogue are built using the virial theorem which results in the following equation for total mass \citep{tempel2014}:
\begin{equation}
  M_{\rm tot} = 2.325 \times 10^{12}\frac{R_g}{\rm Mpc}\left(\frac{\sigma_v}{100\ {\rm km\,s}^{-1}}\right)^{2} M_{\odot}
\end{equation}

where $R_g$ is the gravitational radius and $\sigma_v$ is the 3D velocity dispersion---related to the line-of-sight velocity dispersion ($\sigma_{\rm los}$) via $\sigma_v = \sqrt{3}\sigma_{\rm los}$. Therefore mass can be determined by estimating $R_g$ and $\sigma_{\rm los}$. Estimating $R_g$ is done by equating the definition of potential energy within the virial theorem with:
\begin{equation}
  U = 4\pi G \int\limits_{0}^{R_{\rm out}} \frac{M(r)}{r}\rho(r)r^2 dr,
\end{equation}
from \cite{Binney2008}, where $R_{\rm out}$ is the outer radius, $M(r)$ is the mass contained within a sphere of radius $r$ and $\rho(r)$ is the density profile. Naturally, the choice of $\rho(r)$ will influence the final mass estimate and to this end \cite{tempel2014} provide two: one assuming a Hernquist profile \citep{Hernquist1990}, and the other a Navarro-Frenk-White profile \citep[NFW;][]{Navarro1997}. The NFW case carries a further assumption: the profile shape depends on the concentration, for which \cite{tempel2014} adopt a mass--concentration relation from simulations, so the ratio $R_g/\sigma_{\rm sky}$ is not a constant but varies weakly with mass. We do not re-derive this relation; instead we calibrate $R_g/\sigma_{\rm sky}$ as a function of mass directly against the published \cite{tempel2014} NFW masses, so their concentration assumption is inherited implicitly.

While Nessie provides estimates using either profile, the NFW profile is more popular, and this was what was used in \cite{tempel2017} and subsequently in \cite{driver2022a}. Therefore, we assume an NFW profile mass for the SDSS catalogue in the remainder of this work. 

This method has the advantage of not relying on tuning to a particular mock catalogue but does introduce some ambiguity in both the choice of density profile, and subsequent parameters. Notably, either choice of profile still requires a measurement of $\sigma_{\rm los}$, and while this is conceptually identical to $\sigma_{\rm FoF}$ in \cite{robotham2011}, \cite{tempel2017} choose instead to calculate the velocity dispersion using the standard root-mean-squared method for determining velocity dispersion 

\begin{equation}
  \sigma_{\rm los}^2 = \frac{1}{(1 + z_{\rm mean})^2(n-1)}\sum\limits_{i=1}^{n}\left(v_{i} - v_{\rm mean}\right)^2,
\end{equation}
where the mean velocity and redshift of the group are represented as $v_{\rm mean}$ and $z_{\rm mean}$ respectively, and $n$ and $v_i$ are the number of members in the group and the velocity of member $i$ respectively. It is worth noting here that switching either velocity dispersion estimate, i.e., substituting $\sigma_{\rm FoF}$ for $\sigma_{\rm los}$ and vice versa, will naturally result in different mass estimates, even if all other variables are consistent. In fact, even the choice of group centre can be an area of disagreement; \cite{tempel2014} measure the projected size of the group, $\sigma_{\rm sky}$, about the unweighted mean position of the members, whereas the GAMA catalogues prefer the iterative centre of light \citep{robotham2011}, which is less sensitive to outlying members. We adopt the iterative centre throughout, including for the SDSS NFW masses. On groups with identical membership in both catalogues this is the only remaining difference between our SDSS masses and those of \cite{tempel2017}, and it raises them by a median of 0.075~dex. This all goes to highlight the fact that the mass estimates of any given survey can vary greatly because of minor, \textit{justifiable} choices one way or the other. 

We make these explicit mass estimation choices for the HMF but note that the presented group catalogues in this work also contain many different mass estimates.

\subsection{Determining the Halo Mass Function}
We closely follow the methodology described in \S 3 of \cite{driver2022a} but deviate in several notable cases. Namely, calculating the maximum per-galaxy redshift ourselves, applying a mass-to-light ratio cut to remove groups with erroneous weightings in the HMF, constructing the mass-error function for both the Nessie GAMA and SDSS catalogues, including cosmic variance in the fitted uncertainties, normalizing each survey to a common amplitude, and finally, fitting the HMF using \textsc{emcee} \citep{emcee-package}. We highlight these deviations explicitly when they arise.

We adopt the Murray, Robotham, Power \citep[MRP;][]{Murray2018} function as the theoretical $\Lambda$CDM line to which we fit our data. The MRP is defined as:
\begin{equation}
  \phi(\log_{10}(M/M_{\odot})) = \ln \phi_* \beta \left(\frac{M}{M_*}\right)^{\alpha + 1} {\rm exp}\left[-\left(\frac{M}{M_*}\right)^{\beta}\right]
\end{equation}
where $\phi_*$, $M_*$, $\alpha$, and $\beta$ are the number density, characteristic mass, low-mass slope, and high-mass softening power respectively.

\subsubsection{Building the HMF data points}
\begin{figure*}
  \includegraphics[width=\textwidth]{"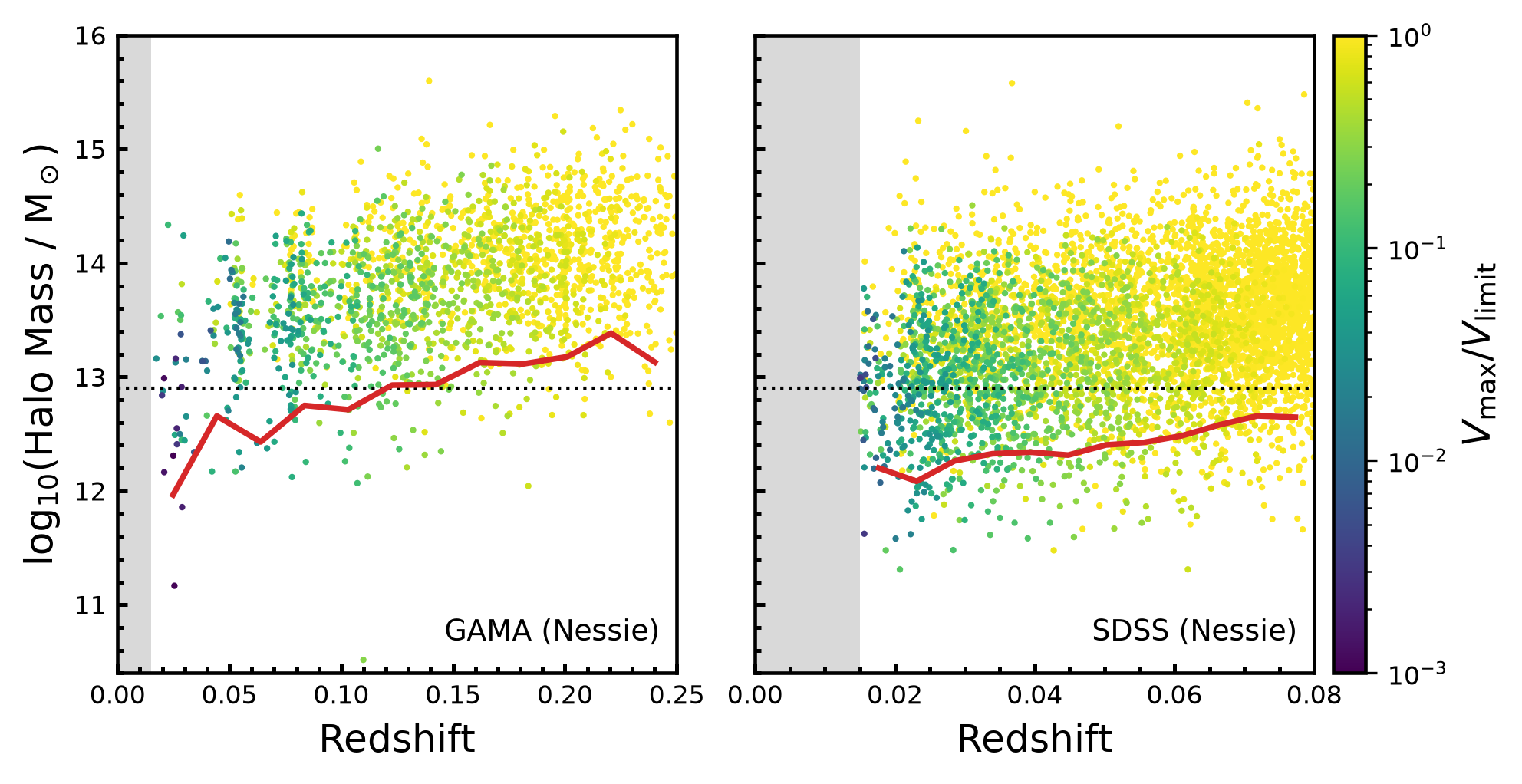"}
  \caption{Halo mass against redshift for the groups that enter the HMF: Nessie GAMA (left; $N\geq5$, $0.015<z<0.25$, after the mass-to-light cut, 1796 groups) and Nessie SDSS (right; $N\geq5$, $0.015<z<0.08$, 4823 groups), with the masses used in the HMF. Each group is coloured by $V_{\rm max}/V_{\rm limit}$, the fraction of the survey volume over which it would remain in the sample; groups at unity are visible throughout, so regions dominated by them are volume limited. The red line is the 5th percentile of $\log_{10}M$ in redshift bins, an empirical lower envelope of the sample. The dotted line marks $\log_{10}(M/M_\odot) = 12.9$, the lowest mass included in the fit (\S\ref{sec:fitting}), and the grey band the excluded region $z<0.015$.}
  \label{fig:z_vs_mass}
\end{figure*}

We build the HMF using the standard $1/V_{\rm max}$ approach (described in more detail below). Each group's $1/V_{\rm max}$ contribution is determined by the maximum observable redshift $z_{\rm max}$ of its $n^{\rm th}$ brightest member, where $n=5$ is the multiplicity cut defining our HMF sample, with $z_{\rm max}$ bounded below by the group redshift and above by the survey limit. We adopt the same redshift windows as \cite{driver2022a}: $0.015 < z < 0.25$ for GAMA and $0.015 < z < 0.08$ for SDSS. The lower limit excludes the very nearest groups, whose distances are most affected by peculiar velocities and whose $V_{\rm max}$ (and hence weight) would be extreme. The upper limits reflect the depth of each survey: at $r<19.65$ (GAMA) and $r<17.77$ (SDSS) an $N\geq5$ group beyond these redshifts must be very massive to have five members above the flux limit, so further volume adds few groups while widening the range of masses that are only visible in part of the survey. Figure~\ref{fig:z_vs_mass} shows the resulting HMF samples in the redshift--mass plane. SDSS is close to volume limited above $\log_{10}(M/M_\odot) \approx 13.5$, where the median group is visible throughout the full volume. For GAMA this is only true above $\approx 14.3$; below this the $1/V_{\rm max}$ weights increasingly come from nearby groups, and the lower envelope of the sample rises from $\log_{10}(M/M_\odot) \approx 12$ at $z \approx 0.02$ to $\approx 13.3$ at $z \approx 0.22$.

The first deviation from \cite{driver2022a} is in how $z_{\rm max}$ is obtained. It was given in the original GAMA \textsc{StellarMassesv19} DMU which \cite{driver2022a} used, but for an $r<19.8$ selection, and does not exist for the new $19.65$ selection. So we calculate $z_{\rm max}$ for each galaxy by solving $M_r + DM(z) + (k+e)(z) = 19.65$, where $M_r$ is the absolute magnitude, $DM(z)$ is the distance modulus at $z$ in the cosmology of the group catalogue, and we calculate the $(k+e)$ using the same method described in \cite{robotham2011}. Explicitly:
\begin{equation}
  (k+e)(z) = \sum\limits_{i=0}^{N} a_i (z_{\rm ref}, z_p)(z-z_p)^i - Q_{z_{\rm ref}} (z - z_{\rm ref})
\end{equation}

where $z_{\rm ref}$ is the redshift that galaxies are $(k+e)$ corrected to, and is set to 0, $Q_{z_{\rm ref}}$ is a luminosity evolution parameter which, at $z=0$, is $Q_0 = 1.75$, $z_p$ is a reference redshift set to 0.2, $a_i(z_{\rm ref}, z_p)$ is the $i^{th}$ coefficient of the polynomial fit to the median $k$-corrections from \textsc{KCORRECTv4.2} \citep{Blanton2007} for $N=4$, with $\vec{a} = [0.2085,1.0226,0.5237,3.5902, 2.3843]$.

The second deviation from \cite{driver2022a} comes from adopting a mass-to-light cut which removes, before binning, any group whose $\log_{10}(M/L_r)$ exceeds the running median at the same mass by more than 0.8~dex. This threshold lies well beyond the 97.5th percentile of the excess ($+0.6$~dex), so the cut is quite liberal, only affecting 11 out of 1807 groups (at $<0.25$). All of these are low-multiplicity ($N \le 9$) systems at low redshift ($z < 0.08$) whose masses are far too large for their luminosity, either because of unrealistic velocity dispersions or because they are unusually faint. Their low redshifts also give them very small $V_{\rm max}$ values ($<4$ per cent of the survey volume), and the combination of the two results in an extreme weighting.

In order to correct for the Eddington bias, we then follow \cite{driver2022a} exactly by reassigning the mass of each group within its lognormal error distribution (\S\ref{sec:mass-error}) and rebuilding the HMF 1001 times. The ratio of the original HMF and the mean of these realizations results in the multiplicative Eddington bias correction within each bin. This correction is then applied to give the debiased HMF, where the uncertainty for each bin is derived from the scatter of the 1001 realizations. We note that this correction is approximate. The observed masses already contain the measurement scatter, so perturbing them again convolves an already-convolved distribution, and the symmetric lognormal kernel does not capture the asymmetry of the mass errors seen in Fig.~\ref{fig:vuvuzela}. The correction is therefore of roughly the right size but not exact, particularly where the HMF is steep. A rigorous treatment requires forward modelling the observed masses within a Bayesian hierarchical fit, which in turn requires a mock catalogue that faithfully reproduces the group-finding efficiency of the data; we leave this to future work. Poisson errors are taken as root-$n$ on the raw counts and the cosmic variance from equation~2 of \citet{Driver2010}, using the four GAMA regions for GAMA and 5 per cent for REFLEX II, as in \citet{driver2022a}. In a further deviation, the three terms are combined in quadrature to give the total error on each point; \citet{driver2022a} instead only include cosmic variance when perturbing the data in their Monte-Carlo refits, which we do not use (\S\ref{sec:fitting}).

\subsubsection{Mass-error functions for SDSS and GAMA}
\label{sec:mass-error}

The Eddington bias correction in \cite{driver2022a} requires an estimate of the mass error as a function of multiplicity. This mass-error function is calculated by taking groups with multiplicity $>20$, iteratively removing the faintest member (in apparent magnitude), recalculating the mass of the group ($M_n$) with the same mass estimator used for the HMF, and comparing this to the original mass with all its members ($M_{\rm max}$). For every group this builds a track of multiplicity vs $\log_{10} (M_n/M_{\rm max})$. From these tracks it is then trivial to determine the 16, 50, and 84 percentiles. Half the 16--84 range gives the mass error, with a floor of 0.1~dex, and for GAMA the median is also used to correct the masses for their multiplicity bias.

\begin{figure}
    \centering
    \includegraphics[width=\linewidth]{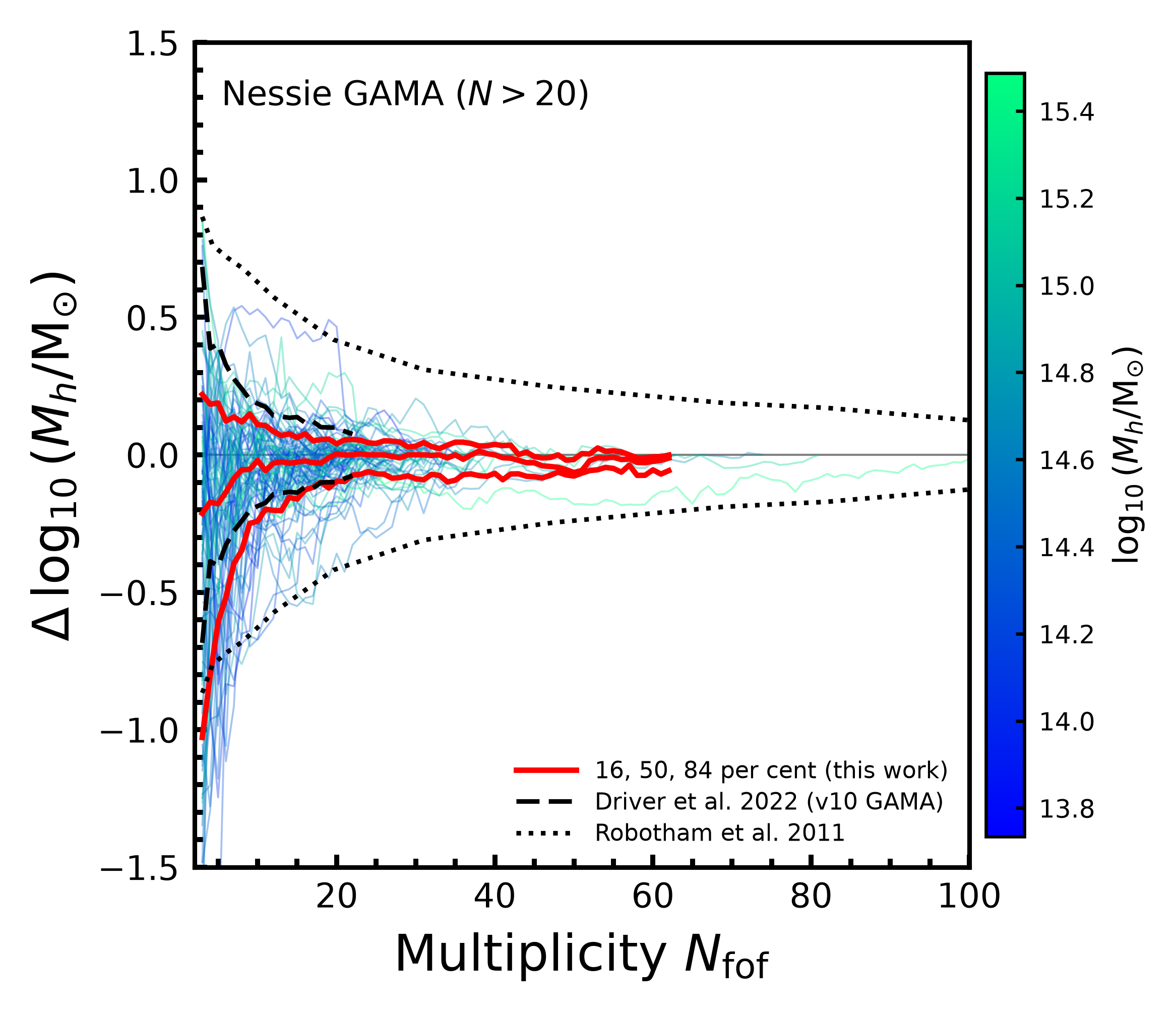}
    \includegraphics[width=\linewidth]{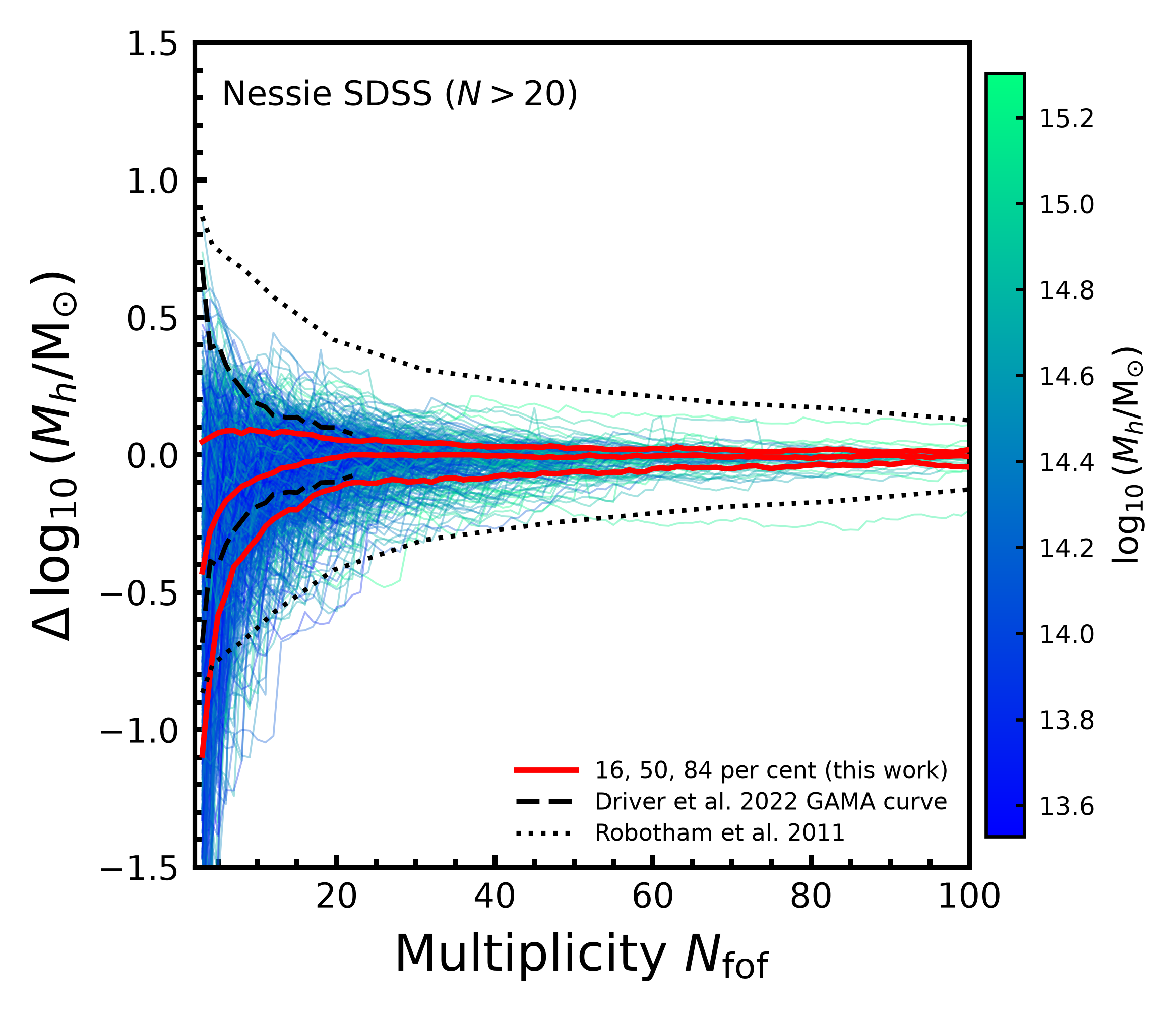}
    \caption{Mass-error functions for both the GAMA (top) and SDSS (bottom) group catalogues. The dotted line shows the mass-error estimates in \citet{robotham2011}. The dashed line highlights the results determined in \citet{driver2022a} which was only run on GAMA. The red-solid lines are the 16, 50, and 84 percentiles of the current group catalogue. The individual tracks for groups with multiplicity $>20$ are shown up to multiplicity of 100, coloured by halo mass.}
    \label{fig:vuvuzela}
\end{figure}

Figure \ref{fig:vuvuzela} shows the mass-error function for the GAMA (top) and SDSS (bottom) groups. In total our GAMA and SDSS catalogues have 78 and 677 groups with multiplicity $>20$ respectively.

The scatter of the two mass-error functions agrees to within $\approx0.05$~dex at all multiplicities, and with that found by \citet{driver2022a}. This is expected, since at low multiplicity it is dominated by the uncertainty in the velocity dispersion, to which both estimators are equally sensitive ($M \propto \sigma^2 R$). The median offsets differ, however: the SDSS masses decline more steeply and more asymmetrically as members are removed. We attribute this to the radius term. The GAMA estimator uses $R_{50}$, which is insensitive to the loss of individual members, whereas the NFW estimator uses the rms projected radius, which is dominated by the outer, typically fainter, members that are removed first.

\subsubsection{Normalizing to SDSS}
In our fit of the HMF we treat $\phi^*$ as a nuisance parameter and choose to normalize each survey to the Nessie SDSS (which has the largest area). $\phi^*$ is still sampled and marginalized over in the fit. This allows us to better compare the resulting shape of the HMF from the different data sets, albeit not the actual number densities.

We take the mass range $14 < \log_{10}(M/M_\odot) < 15$, where all surveys overlap with one another and have an adequate number of counts. For every survey we integrate the binned number density over this range, interpolating linearly in $\log_{10}\phi$ between bins. Each survey's $\phi$ is then multiplied by the single constant that brings its integral onto that of the Nessie SDSS catalogue; the corresponding effective volume is divided by the same constant, and the fractional errors are unaffected. The MRP is fitted to the rescaled points so that it carries only the shape $(M^{*}, \alpha, \beta)$ together with one common, arbitrarily chosen amplitude. Because a difference in mass calibration shifts the HMF horizontally rather than vertically, this removes the mean offset between surveys but not any mass dependence.

\subsubsection{Fitting the HMF}
\label{sec:fitting}
The method for fitting the HMF to the datapoints is the final deviation from \cite{driver2022a}. We fit the MRP function by the same criterion as \citet{driver2022a}, their equation~(6): a $\chi^2$ over the occupied bins together with a penalty term
\begin{equation}
2\,\Delta\log_{10}M \sum_i \bar\phi_i \sum_j V_j
\end{equation}
evaluated over the ten 0.2~dex bins above the highest-mass bin in the fit, where $V_j$ is the (rescaled) volume of survey $j$. Only bins above $\log_{10}(M/M_\odot) = 12.9$ enter the fit for GAMA and SDSS, while all REFLEX II bins are used. We note that this penalty is exactly $-2\ln P(0|\mu_i)$ for Poisson counts with expectation $\mu_i = \bar\phi_i \sum_j V_j\,\Delta\log_{10}M$, and the $\chi^2$ is Gaussian in $\log_{10}\phi$, so their equation~(6) is already a valid $-2\ln\mathcal{L}$. We therefore adopt $\ln\mathcal{L} = -\tfrac{1}{2}\chi^2$ without modification, and sample the posterior with \textsc{emcee} \citep{emcee-package} under flat priors on $\log_{10}M^* \in [11, 16.5]$, $\log_{10}\phi^* \in [-8, 0]$, $\alpha \in [-2.5, 0.5]$ and $\beta \in [0.05, 3]$, using 64 walkers for 25\,000 steps and discarding the first 6\,250. Parameter uncertainties follow from the posterior, in place of the perturb-and-refit Monte Carlo of \citet{driver2022a}. Apart from the inclusion of cosmic variance in the errors, the objective function is thus identical to theirs, and applied to their own data our sampler recovers their best fit ($\chi^2 = 211.55$ in both), so any difference between our fitted parameters and theirs is attributable to the data rather than to how the surface is explored.

\section{Results}
\subsection{Group Catalogues}


\begin{table*}
\caption{Table comparing the number of groups between the previous GAMA Group Catalogue and this work.}
\begin{tabular}{lrrrrrrrr}
\hline
 & \multicolumn{4}{c}{$G^{3}C$v$_{10}$} & \multicolumn{4}{c}{This work} \\
Field & $N \geq 2$ & $N \geq 3$ & $N \geq 5$ & $N \geq 20$ & $N \geq 2$ & $N \geq 3$ & $N \geq 5$ & $N \geq 20$ \\
\hline
G02 & 2\,540 & 940 & 307 & 16 & -- & -- & -- & -- \\
G09 & 7\,481 & 2\,807 & 883 & 48 & 6\,181 & 2\,141 & 582 & 15 \\
G12 & 8\,171 & 3\,002 & 955 & 68 & 7\,028 & 2\,378 & 667 & 27 \\
G15 & 8\,002 & 2\,969 & 916 & 61 & 6\,964 & 2\,427 & 668 & 23 \\
G23 & -- & -- & -- & -- & 5\,144 & 1\,763 & 475 & 18 \\
\hline
G09+G12+G15 & 23\,654 & 8\,778 & 2\,754 & 177 & 20\,173 & 6\,946 & 1\,917 & 65 \\
All fields & 26\,194 & 9\,718 & 3\,061 & 193 & 25\,317 & 8\,709 & 2\,392 & 83 \\
\hline
\end{tabular}
\label{tab:group_comp}
\end{table*}

\begin{figure*}
  \includegraphics[width = 0.45\textwidth]{"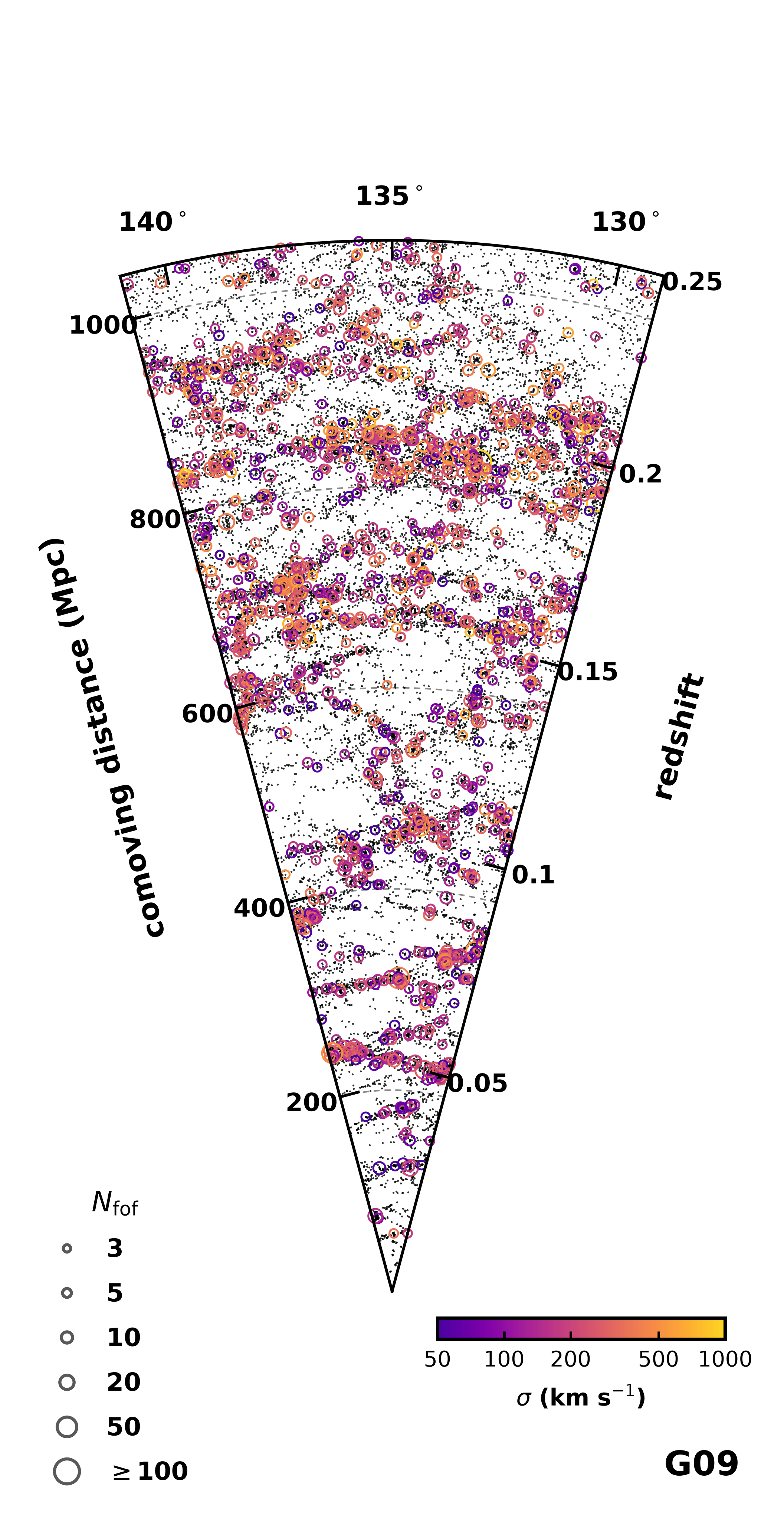"}
  \includegraphics[width = 0.45\textwidth]{"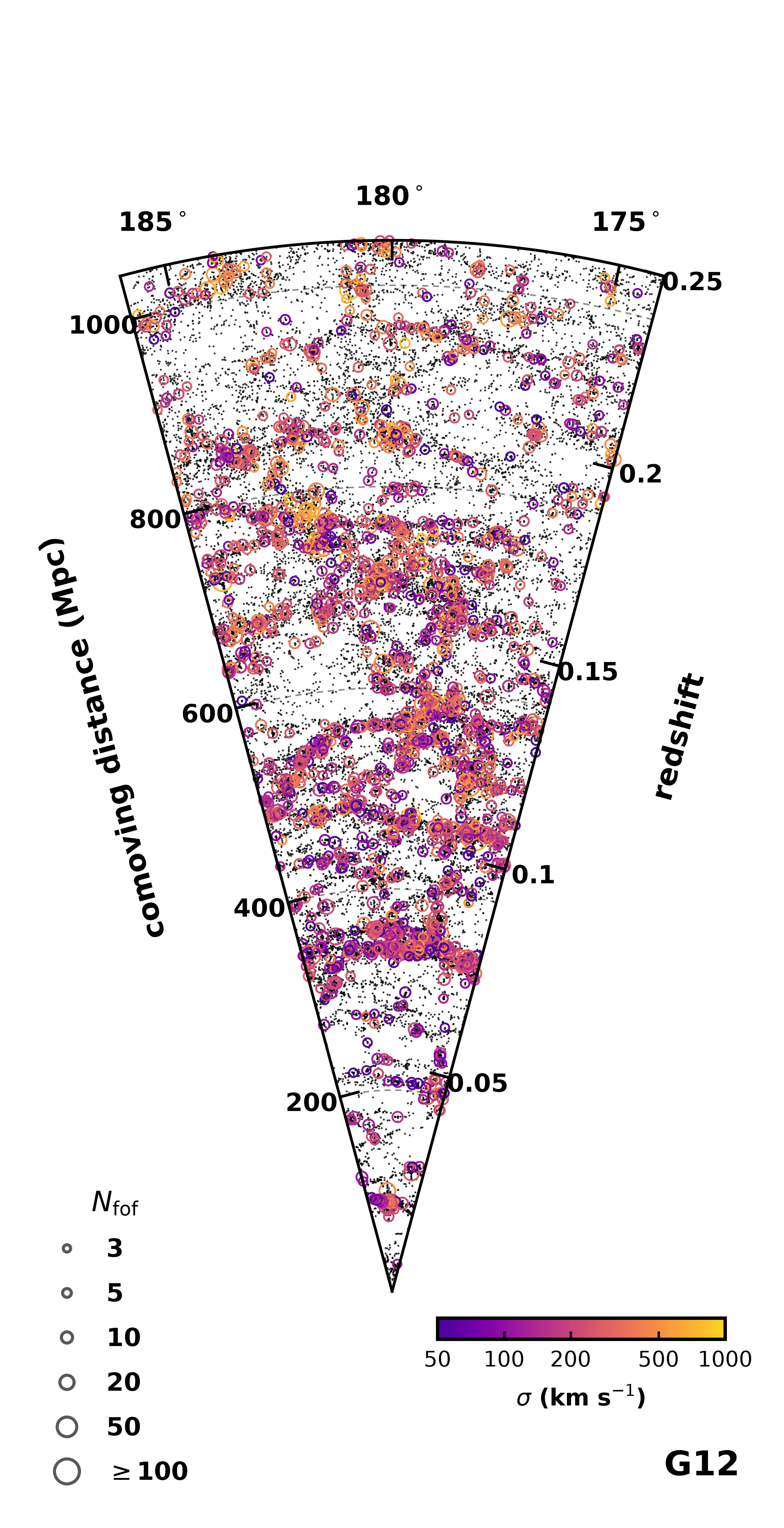"}
  \caption{Wedge diagrams of the GAMA group catalogue for: G09 and G12. Galaxies are shown as small black points and groups with $N\geq3$ as circles, sized by multiplicity and coloured by velocity dispersion.}
  \label{fig: wedge_g09g12}
\end{figure*}

\begin{figure*}
  \includegraphics[width = 0.45\textwidth]{"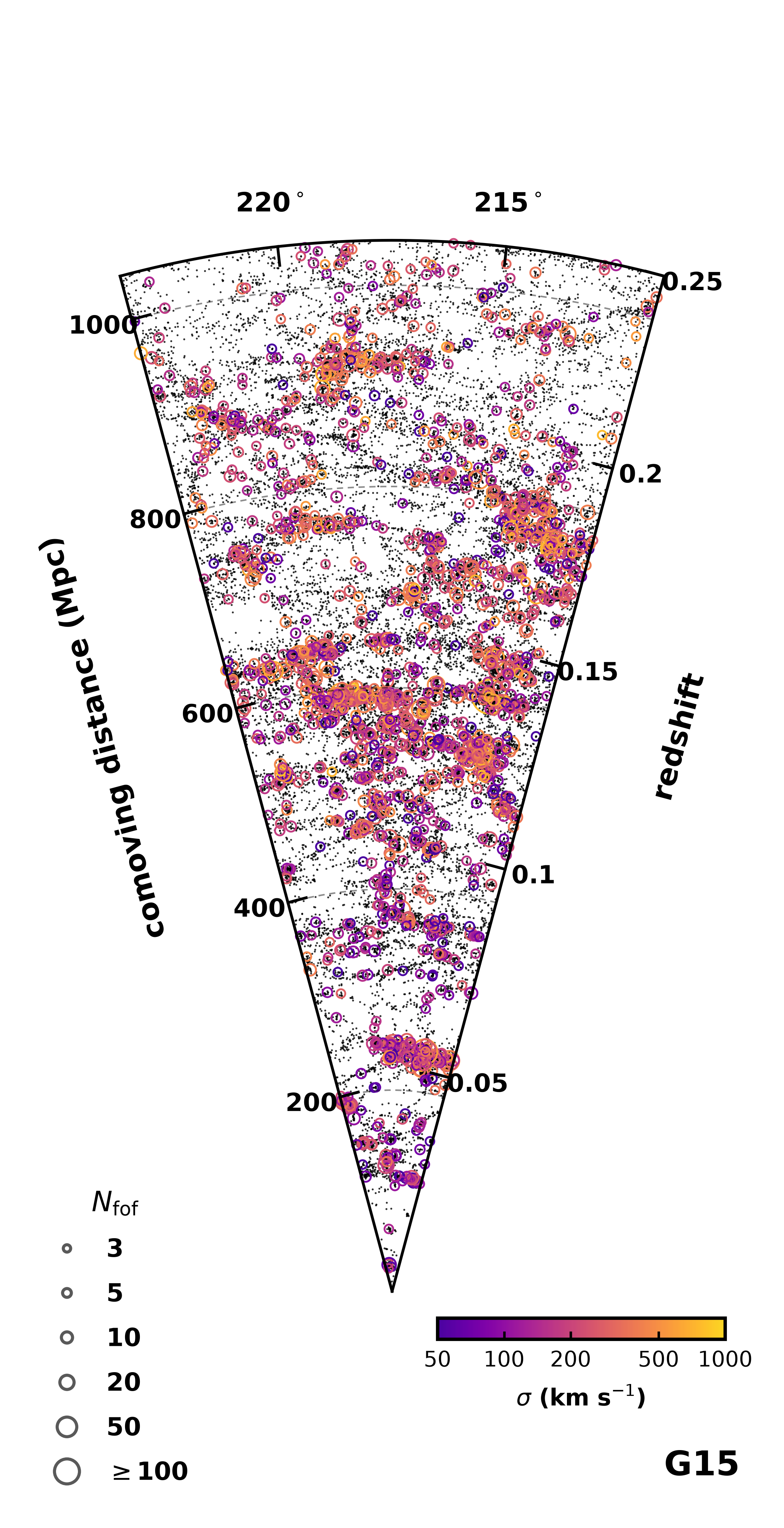"}
  \includegraphics[width = 0.45\textwidth]{"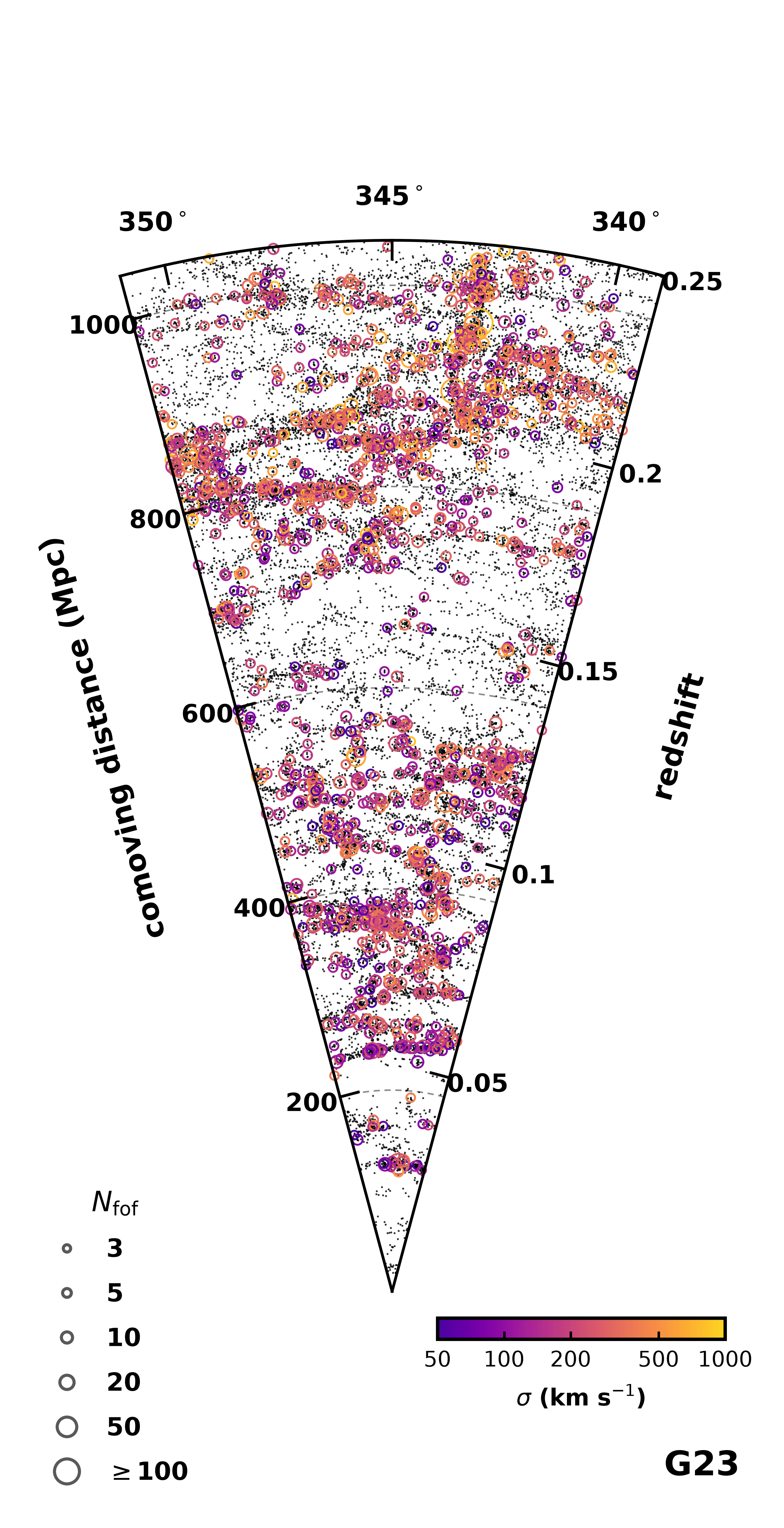"}
  \contcaption{Wedge diagrams of the GAMA group catalogue for: G15 and G23. Galaxies are shown as small black points and groups with $N\geq3$ as circles, sized by multiplicity and coloured by velocity dispersion.}
  \label{fig: wedge_g15g23}
\end{figure*}

\cite{lambert2026} showed that the Nessie results, when applied to GAMA data in particular, are equivalent to the group finder used in \cite{robotham2011}, and indeed, one of the unit tests is to exactly recover their results. This means that any differences in the catalogue are driven by the updated photometry as opposed to any methodology changes between the group finders.

The resultant wedge plots are shown in Fig. \ref{fig: wedge_g09g12}. The black points in the background are the latest GAMA release galaxy points from \cite{driver2022} with the improved photometry from \cite{bellstedt2020}. The circles are the $G^{3}C$v$_{11}$ groups, sized by number of members and coloured by velocity dispersion, following \cite{robotham2011}. Since mass scales as $\sigma^2$, the colours give a visual impression of the halo masses the catalogue reaches at each distance: nearby, groups with low dispersions are recovered, whereas at higher redshift the catalogue is increasingly dominated by high-dispersion systems. We only show the galaxies up to $z=0.25$ (since this is the sample that was used in the HMF) but the full catalogue is cut at $z=0.5$.

The group number counts across the different fields for varying multiplicity cuts are presented in Table \ref{tab:group_comp} for both $G^{3}C$v$_{11}$ and $G^{3}C$v$_{10}$.  

\begin{figure*}
  \includegraphics[width = \textwidth]{"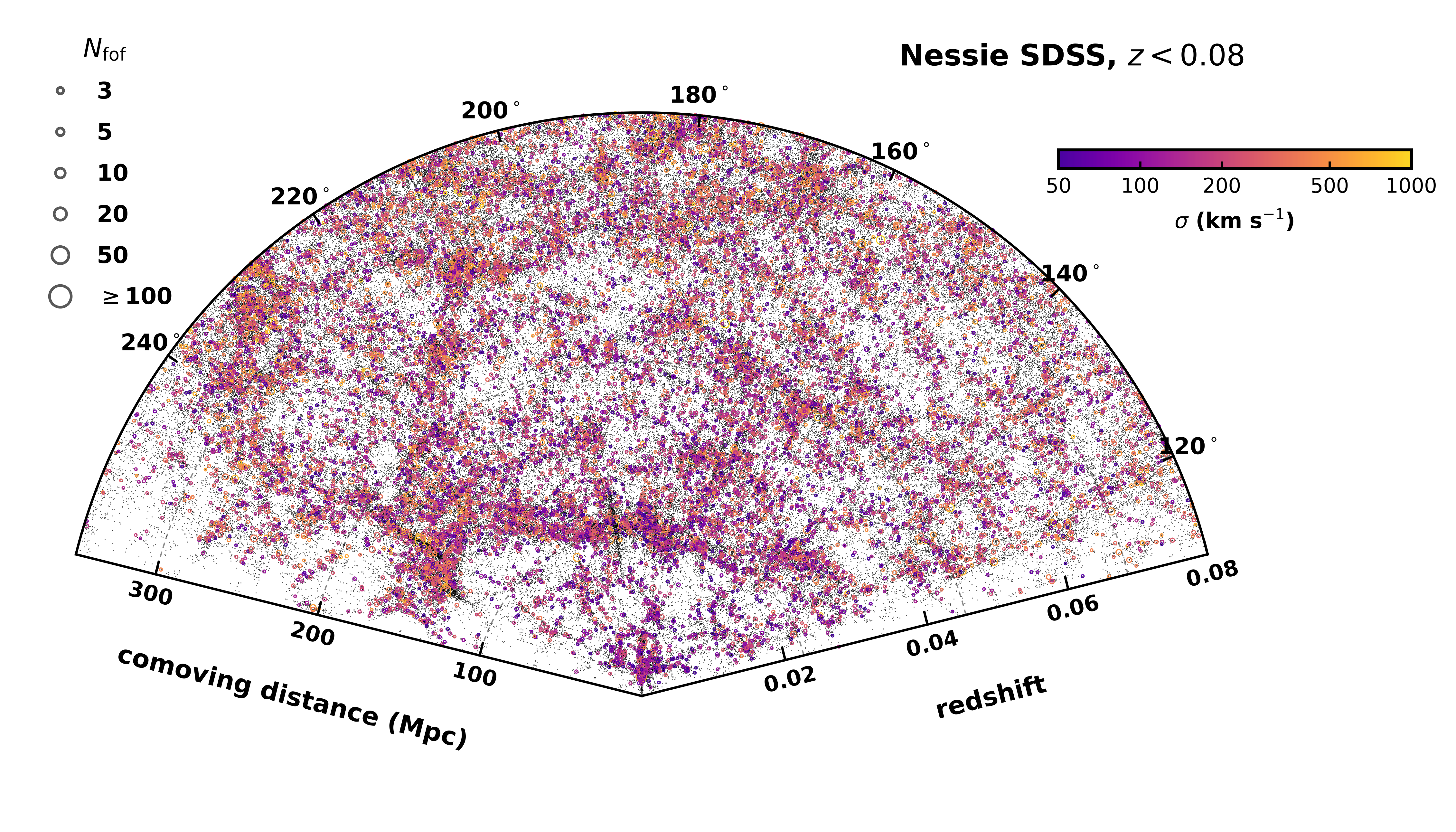"}
  \caption{Wedge diagram of the SDSS group catalogue. The circles highlight groups with $N\geq3$, sized by multiplicity and coloured by velocity dispersion, and the galaxies are shown as small black points in the background.}
  \label{fig: sdss-wedge}
\end{figure*}

We also include the group catalogue that we built from the SDSS sample as a data product. Figure \ref{fig: sdss-wedge} is the wedge plot, highlighting the SDSS galaxies (black small points) that were used to construct the group catalogue (circles, coloured by velocity dispersion as in Fig.~\ref{fig: wedge_g09g12}). Table \ref{tab:sdss_comparison} presents the group number counts in varying redshift and multiplicity bins for both this work and \cite{tempel2017}.

\begin{table*}
\caption{Table comparing the number of groups in redshift ranges and multiplicities between \citet{tempel2017} and this work.}

\begin{tabular}{lrrrrrrrr}
\hline
 & \multicolumn{4}{c}{Tempel et al. (2017)} & \multicolumn{4}{c}{This work} \\
Redshift & $N \geq 2$ & $N \geq 3$ & $N \geq 5$ & $N \geq 20$ & $N \geq 2$ & $N \geq 3$ & $N \geq 5$ & $N \geq 20$ \\
\hline
$0.00 \leq z < 0.02$ & 1\,544 & 755 & 258 & 18 & 1\,536 & 664 & 246 & 17 \\
$0.02 \leq z < 0.04$ & 7\,705 & 3\,808 & 1\,262 & 131 & 7\,086 & 3\,231 & 1\,262 & 133 \\
$0.04 \leq z < 0.06$ & 9\,991 & 4\,663 & 1\,392 & 111 & 9\,012 & 3\,759 & 1\,349 & 122 \\
$0.06 \leq z < 0.08$ & 15\,293 & 7\,071 & 2\,104 & 160 & 13\,772 & 5\,784 & 2\,129 & 183 \\
\hline
$0.015 < z < 0.08$ & 33\,615 & 15\,840 & 4\,847 & 408 & 30\,442 & 13\,024 & 4\,823 & 445 \\
All groups & 88\,662 & 37\,365 & 10\,087 & 642 & 84\,888 & 33\,502 & 11\,161 & 742 \\
\hline
\end{tabular}
\label{tab:sdss_comparison}
\end{table*}

One of Nessie's main objectives is to be as generalized as possible and as such we do not recover all the properties that were calculated in previous versions of the GAMA group catalogue since these were either 1) specific to GAMA, 2) relied on particular assumptions that we are not comfortable making for all group catalogues, or 3) were not widely used within the community. These legacy properties can all be calculated from the base catalogue if necessary, and in addition, more group properties are constantly being added to Nessie and updating the group properties using Nessie is a straight-forward exercise.

We retire the following properties in this version of the group catalogue: RelDen, Zcomp, GroupEdge which are dependent on the randoms catalogue used; LinkTot, LinkStrenProj, LinkStrenRad, LinkStrenComb, which are all the linking statistics that Nessie does not provide; RadKurt, AxAng, AxRat, SymKS, VelSkew, VelKurt, Modality, d3vol, d3area, d2radec, d2raz, d2decz, distortcirc, distortellip, distortsph, which are projection and geometrical measurements which can be calculated in a variety of ways; VelDispRaw, Rgap, multiple steps in the gapper velocity dispersion measurement; IterCenCATAID, BCGCATAID where we do not keep track of the CATAIDs in Nessie; and URL\_IMG, since these were not available. In practice these outputs saw very little use over the roughly 15 years since the original catalogue was released, so we determined that they were not worth preserving.

We provide a full catalogue of the properties that are available in $G^{3}C$v$_{11}$ and the Nessie SDSS group catalogue in Table \ref{tab: properties} and explicitly note when some properties are available in one and not the other. 

\begin{table*}
  \caption{List of group properties that are provided in $G^{3}C$v$_{11}$ and the Nessie SDSS group catalogues and the differences between the two.}
\begin{tabular}{llcccp{7.5cm}}
\hline
Column & Units & v10 & GAMA & SDSS & Description \\
\hline
GroupID & -- & $\checkmark$ & $\checkmark$ & $\checkmark$ & Unique group ID, as listed in the associated galaxy catalogue. For GAMA this is the Nessie ID plus a region offset (1e5 G09, 2e5 G12, 3e5 G15, 5e5 G23) \\
Nfof & -- & $\checkmark$ & $\checkmark$ & $\checkmark$ & Group multiplicity \\
IterCenID & -- & IterCenCATAID & $\checkmark$ & $\checkmark$ & ID of the iterative central galaxy (preferred position of the group centre): GAMA III UberID for GAMA, \citet{tempel2014} galid for SDSS. Not a CATAID \\
IterCenRA & deg & $\checkmark$ & $\checkmark$ & $\checkmark$ & RA of the iterative central galaxy (J2000) \\
IterCenDec & deg & $\checkmark$ & $\checkmark$ & $\checkmark$ & Dec of the iterative central galaxy (J2000) \\
IterCenZ & -- & $\checkmark$ & $\checkmark$ & $\checkmark$ & Redshift of the iterative central galaxy \\
Zfof & -- & $\checkmark$ & $\checkmark$ & $\checkmark$ & Median redshift of the group \\
Rad50 & $h^{-1}$\,Mpc & $\checkmark$ & $\checkmark$ & $\checkmark$ & Group radius defined by the 50th percentile group member, based on the projected comoving distance from IterCen \\
Rad1Sig & $h^{-1}$\,Mpc & $\checkmark$ & $\checkmark$ & $\checkmark$ & Group radius defined by the 68th percentile group member, based on the projected comoving distance from IterCen \\
Rad100 & $h^{-1}$\,Mpc & $\checkmark$ & $\checkmark$ & $\checkmark$ & Group radius defined by the most distant group member, based on the projected comoving distance from IterCen \\
VelDisp & km\,s$^{-1}$ & $\checkmark$ & $\checkmark$ & $\checkmark$ & Gapper velocity dispersion with VelErr subtracted in quadrature; set to zero if VelErr exceeds the raw dispersion \\
VelErr & km\,s$^{-1}$ & $\checkmark$ & $\checkmark$ & $\checkmark$ & Total group velocity error subtracted from VelDisp (input galaxy velocity errors of 50\,km\,s$^{-1}$) \\
TotRmag & mag\,$-\,5\log h$ & $\checkmark$ & $\checkmark$ & $\checkmark$ & $(k+e)$-corrected $r$-band absolute magnitude of the group, from the summed member luminosities (GAMA: DMU $(k+e)$ polynomial; SDSS: \citet{tempel2014} KCORRECT) \\
TotFluxProxy & -- & $\checkmark$ & $\checkmark$ & $\checkmark$ & Proxy for group total $r$-band luminosity down to $M_r - 5\log h = -14$ in solar luminosities, from the observed flux and the fraction of the GAMA LF observable at the group redshift ($B = 1$ in equation~22 of \citealt{robotham2011}). SDSS uses the $r < 17.77$ limit \\
MassProxy & $h^{-1}\,\mathrm{M}_\odot$ & $\checkmark$ & $\checkmark$ & $\checkmark$ & Dynamical mass proxy Rad50 times VelDisp squared over $G$, i.e. $A = 1$ in equation~18 of \citet{robotham2011} \\
BCGID & -- & BCGCATAID & $\checkmark$ & $\checkmark$ & ID of the brightest group galaxy: GAMA III UberID for GAMA, \citet{tempel2014} galid for SDSS. Not a CATAID \\
BCGRA & deg & $\checkmark$ & $\checkmark$ & $\checkmark$ & RA of BCG (J2000) \\
BCGDec & deg & $\checkmark$ & $\checkmark$ & $\checkmark$ & Dec of BCG (J2000) \\
CenRA & deg & $\checkmark$ & $\checkmark$ & $\checkmark$ & RA of the $r$-band luminosity weighted centre of the group (J2000) \\
CenDec & deg & $\checkmark$ & $\checkmark$ & $\checkmark$ & Dec of the $r$-band luminosity weighted centre of the group (J2000) \\
MassA & $h^{-1}\,\mathrm{M}_\odot$ & $\checkmark$ & $\checkmark$ & -- & MassProxy times the global $A$ factor, $A = 10.0$ (\citealt{robotham2011}, section 4.3). Calibrated on GAMA mocks: GAMA only \\
MassAfunc & $h^{-1}\,\mathrm{M}_\odot$ & $\checkmark$ & $\checkmark$ & -- & MassProxy times the functional $A$ factor, a function of Nfof and Zfof (\citealt{robotham2011}, section 4.3). Calibrated on GAMA mocks: GAMA only \\
LumB & $h^{-2}\,\mathrm{L}_\odot$ & $\checkmark$ & $\checkmark$ & -- & TotFluxProxy times the global $B$ factor, $B = 1.04$ (\citealt{robotham2011}, section 4.4). Calibrated on GAMA mocks: GAMA only \\
LumBfunc & $h^{-2}\,\mathrm{L}_\odot$ & $\checkmark$ & $\checkmark$ & -- & TotFluxProxy times the functional $B$ factor, a function of Nfof and Zfof (\citealt{robotham2011}, section 4.4). Calibrated on GAMA mocks: GAMA only \\
BCGZ & -- & -- & $\checkmark$ & $\checkmark$ & Redshift of BCG \\
MassHern & $h^{-1}\,\mathrm{M}_\odot$ & -- & $\checkmark$ & $\checkmark$ & Dynamical mass from equation~8 of \citet{tempel2014} for a Hernquist profile: $R_g = 4.582\,\sigma_{\rm sky}$, $\sigma_{\rm 3D} = \sqrt{3}\,$VelDisp, with $\sigma_{\rm sky}$ the projected rms member distance from IterCen. Zero where VelDisp is zero \\
MassNFW & $h^{-1}\,\mathrm{M}_\odot$ & -- & $\checkmark$ & $\checkmark$ & Dynamical mass from equation~8 of \citet{tempel2014} for an NFW profile: $R_g = \kappa(M)\,\sigma_{\rm sky}$ with $\kappa$ calibrated on the \citet{tempel2014} NFW masses, $\sigma_{\rm sky}$ the projected rms member distance from IterCen, $\sigma_{\rm 3D} = \sqrt{3}$ times the rms dispersion of their equation~3. Zero where all members share one redshift \\
Region & -- & -- & $\checkmark$ & $\checkmark$ & Survey region: g09, g12, g15 or g23 for GAMA; SDSS for SDSS \\
\hline
\end{tabular}
\label{tab: properties}
\end{table*}

\subsection{Halo Mass Function}
There are various decisions to be made when building the HMF. Principal among them are which surveys to include, and which mass estimates to use. Any combination of these choices results in different fits for the HMF which is clear from Table 2 in \cite{driver2022a} who fit a wide variety of HMFs varying the included data points, multiplicities, different mass estimates, and different fixed values; the spread in the MRP parameters is enormous.

Since this is an update to \cite{driver2022a} we focus on exploring the effect of updating the GAMA data primarily and exploring using the SDSS data in a consistent way to GAMA. Table \ref{tab:mrp} shows the fits for various HMFs that we have constructed to explore these data sets and highlights the two cases where the fits are unconstrained. First, the GAMA data alone; second, combining this with the Nessie SDSS group catalogue; and finally combining these data sets with the REFLEX II data. 

\begin{table*}
\centering
\caption{MRP fits. Values are posterior medians with 16th/84th percentile credible intervals.  Rows marked $\dagger$ are unconstrained: the $M_*$ posterior increases towards the lower prior limit ($\log_{10}M_* = 11$), so their $M_*$ and $\beta$ reflect the prior range rather than the data.  $\chi^2_\nu = \chi^2/\nu$, with $\nu = N_{\rm bins} - 4$ degrees of freedom.}
\begin{tabular}{lcccccc}
\hline
Sample & log$_{10}$(M$_*$/M$_\odot$) & log$_{10}(\phi_*)$ & $\alpha$ & $\beta$ & $\chi^2$ & $\chi^2_\nu$ ($\nu$) \\
 & & Mpc$^{-3}$ dex$^{-1}$ & & & & \\
\hline
GAMA (Nessie)$\dagger$ & $12.13^{+0.94}_{-0.77}$ & $-3.29^{+0.24}_{-0.35}$ & $-0.53^{+0.34}_{-0.41}$ & $0.34^{+0.09}_{-0.05}$ & 14.2 & 1.42 (10) \\
GAMA + SDSS (Nessie)$\dagger$ & $11.70^{+0.70}_{-0.49}$ & $-2.22^{+0.19}_{-0.22}$ & $-0.96^{+0.16}_{-0.20}$ & $0.28^{+0.04}_{-0.02}$ & 112.2 & 3.03 (37) \\
GAMA + SDSS (Nessie) + REFLEX II & $12.70^{+0.59}_{-0.78}$ & $-2.72^{+0.18}_{-0.36}$ & $-1.03^{+0.28}_{-0.23}$ & $0.38^{+0.07}_{-0.06}$ & 184.7 & 2.31 (80) \\
\hline
Murray+21 $\Lambda$CDM, rescaled & $14.43$ & $-4.53$ & $-1.86$ & $0.71$ & -- & -- \\
\hline
\end{tabular}
\label{tab:mrp}
\end{table*}

\begin{figure}
  \includegraphics[width=0.5\textwidth]{"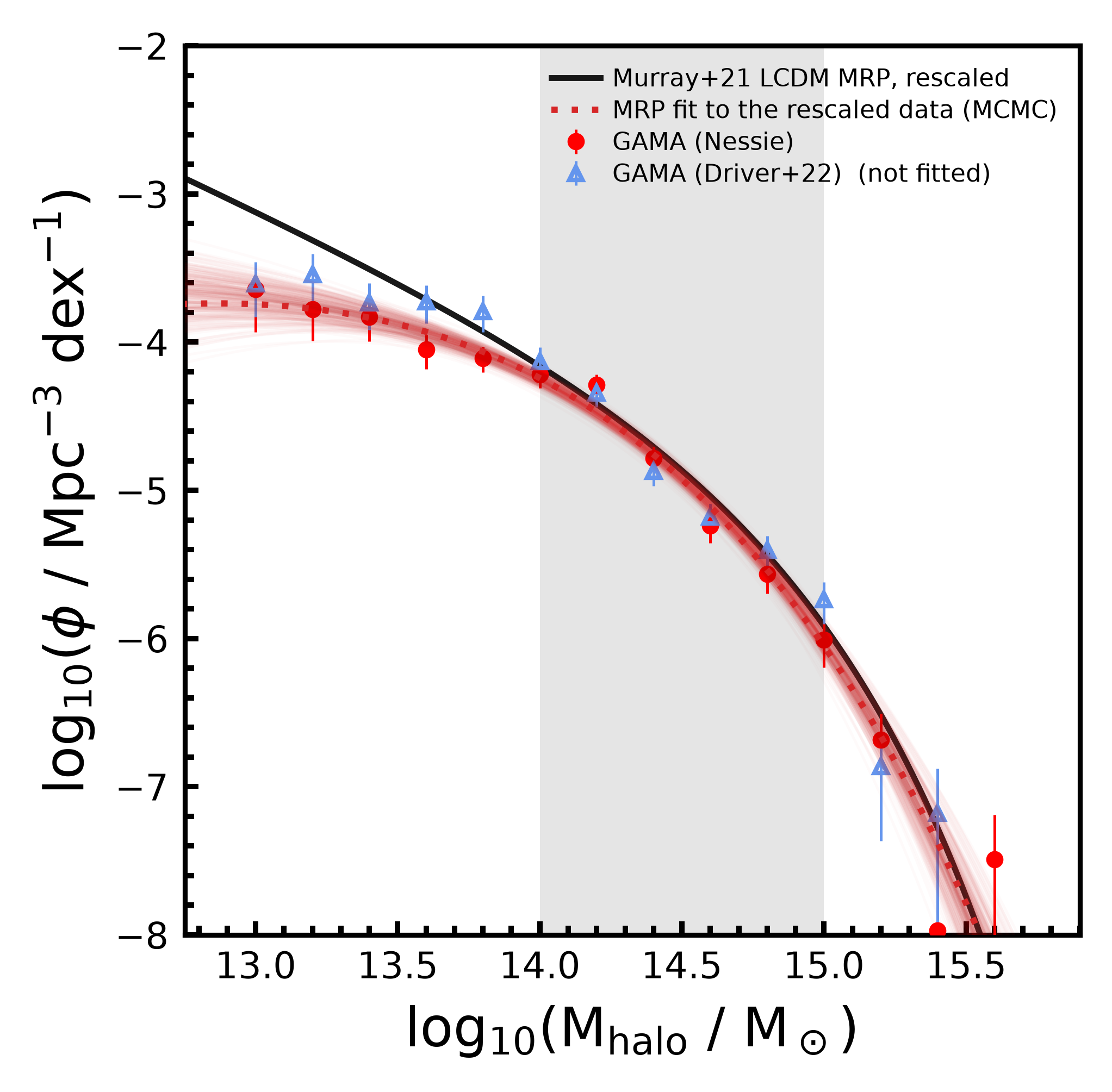"}
  \caption{HMF as a result of just the GAMA data. The GAMA data points from \citet{driver2022a} are shown as blue triangles for comparison, but are not fitted.}
  \label{fig: hmf_gama}
\end{figure}

The first important comparison is to build the HMF just using $G^{3}C$v$_{11}$. Already from \cite{driver2022a} it was known that GAMA was not sufficient on its own in constraining the HMF. This is also what we find in this work (see Table \ref{tab:mrp}) despite the inclusion of a fourth field (G23). Figure \ref{fig: hmf_gama} shows the HMF fitted to $G^{3}C$v$_{11}$. The data points for $G^{3}C$v$_{11}$ are shown in red, the Markov-Chain Monte-Carlo (MCMC) fit to these data points is represented as the red dotted line with MCMC runs included to show the spread in the fit. The expected $\Lambda$CDM MRP from \cite{Murray2021} is shown as the black solid line. As expected, GAMA alone, even with an additional field, is not enough to constrain the HMF. The GAMA-only fit achieves a $\chi^2$ of 14.2 ($\chi^2_\nu = 1.42$ for $\nu = 10$). 

\begin{figure}
  \includegraphics[width=0.5\textwidth]{"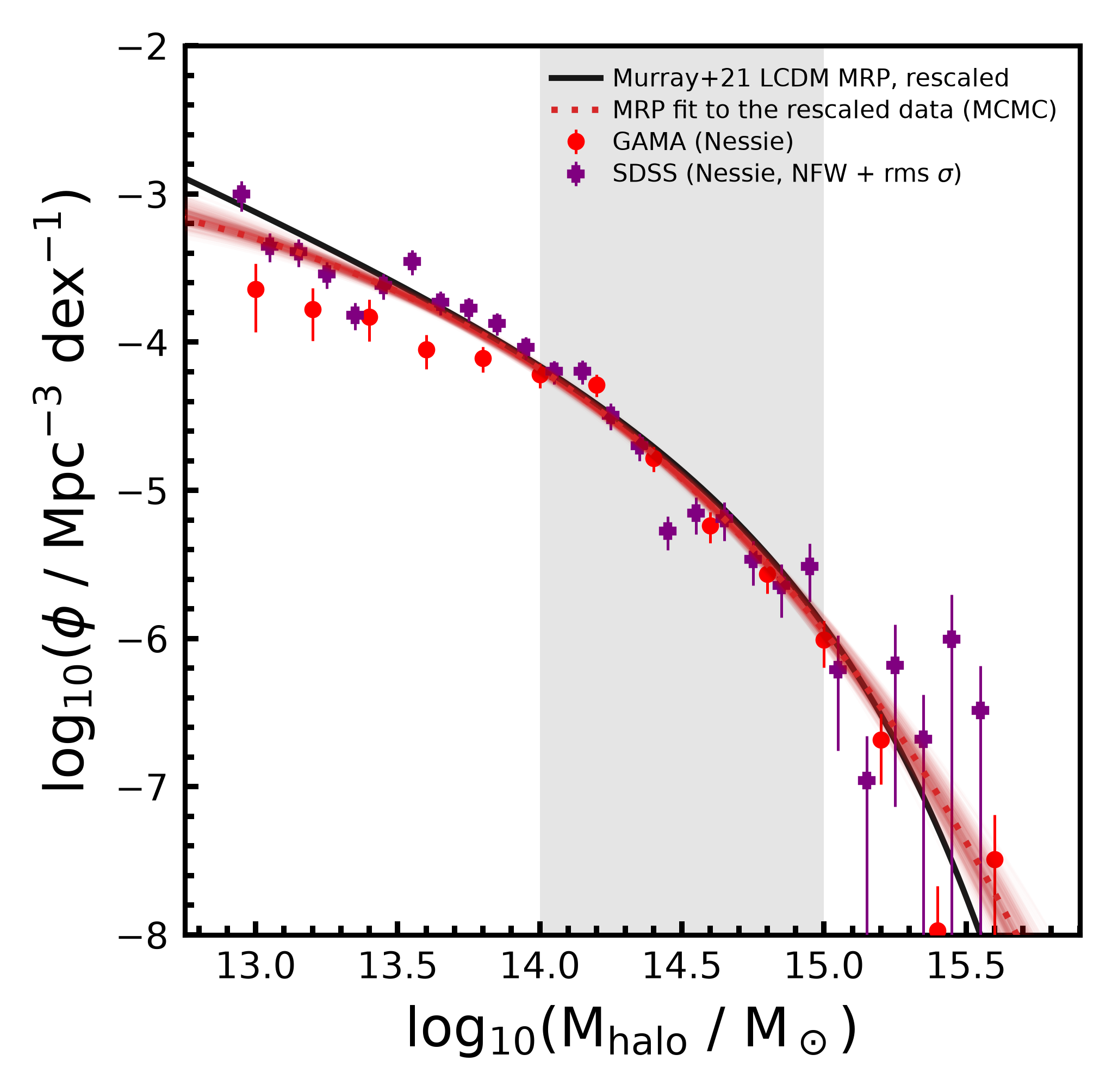"}
  \caption{HMF as a result of just the GAMA and SDSS Nessie data.}
  \label{fig: hmf_gama_sdss}
\end{figure}

The fit from the Nessie SDSS group catalogue combined with the $G^3_{v11}$ is shown in Fig. \ref{fig: hmf_gama_sdss}. Here the points and lines match the meaning of Fig. \ref{fig: hmf_gama} with the inclusion of the SDSS data points in magenta crosses; the red dotted line now shows the fit to both $G^{3}C$v$_{11}$ (red) and SDSS (magenta) points.

This is an improvement over the GAMA-only fit, as to be expected due to the larger volume. With all 4 MRP parameters more tightly constrained, however, the $\chi^2$ fit is now 112.2 ($\chi^2_\nu = 3.03$ for $\nu = 37$) although the fit sits closer to the theoretical value. However, we still see a deviation at low masses. In addition, as for GAMA alone, $M_*$ is unconstrained: its posterior increases towards the lower prior limit, as these data do not reach high enough masses to locate the exponential cut-off. SDSS and GAMA are insufficient to constrain the HMF. This was the same conclusion that \cite{driver2022a} came to and why they include REFLEX II data. We follow the same approach.

\begin{figure*}
  \includegraphics[width=\textwidth]{"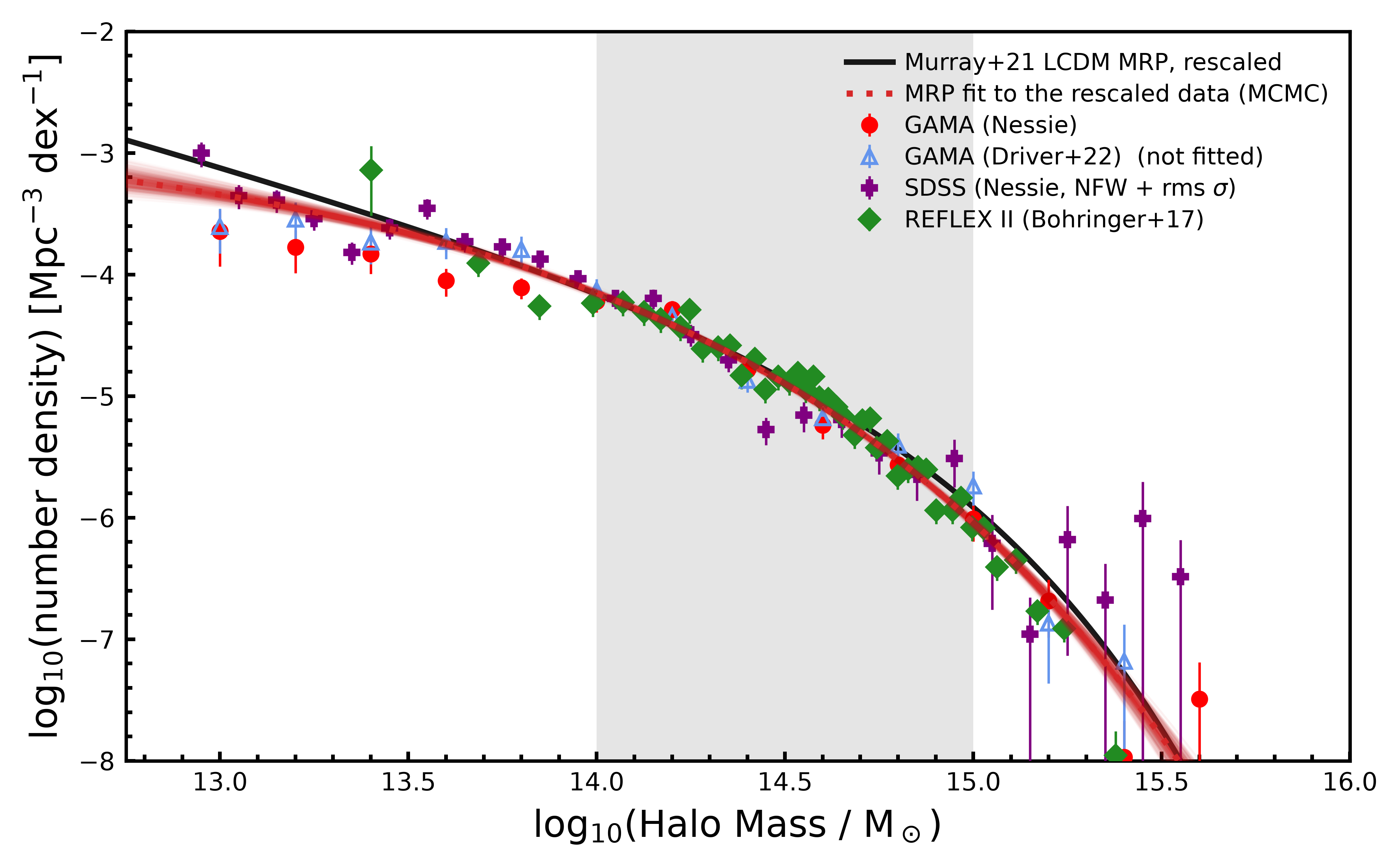"}
  \caption{HMF using the combined Nessie groups for GAMA and SDSS and REFLEX II which are represented as red circles, magenta crosses, and green diamonds respectively. The SDSS groups from \citet{tempel2017} are included for reference as empty orange squares, but are not fitted. The black solid line shows the MRP curve expected from \citet{Murray2021} and the red curve shows our MRP fit to the combined data.}
  \label{fig: hmf_all}
\end{figure*}

Figure \ref{fig: hmf_all} includes the HMF fit with REFLEX II points. It is clear in Fig. \ref{fig: hmf_all} that the fit deviates from the expected $\Lambda$CDM curve at low masses but begins to follow the curve reasonably well at about $\log_{10}(M/M_{\odot}) > 13.5$. This is the only fit whose parameters are constrained by the data, with REFLEX II locating the exponential cut-off ($\chi^2 = 184.7$, $\chi^2_\nu = 2.31$ for $\nu = 80$). The $\chi^2_\nu$ values exceed unity because the surveys still differ in shape after normalisation (SDSS lies 0.3--0.5~dex above GAMA at $10^{13}$--$10^{13.8}\,M_\odot$); the lower $\chi^2_\nu$ of the GAMA-only fit reflects its unconstrained parameters rather than a better fit.

These results closely follow those from \cite{driver2022a} and confirm that Nessie successfully builds a group catalogue for the new GAMA photometry which is similar to the original GAMA catalogue used in \cite{robotham2011}. This is to be expected from the results in \cite{lambert2026} who compare the $n(z)$ between the two catalogues. Further, given that 1) Nessie was built primarily for GAMA, and 2) was built from the original GAMA group finder, the HMF fit being so close to \cite{driver2022a} is a testament to $G^{3}C$v$_{11}$ being on par with previous versions of the GAMA group catalogue.  

\section{Discussion}
\subsection{Comparison between $G^{3}C$v$_{10}$ and $G^{3}C$v$_{11}$}
The most notable difference between $G^{3}C$v$_{10}$ and $G^{3}C$v$_{11}$ is the different fields that exist. There is overlap in the equatorial fields (G09, G12, G15), however v$_{10}$ has groups in G02, even though that field was incomplete, whereas v$_{11}$ does not include any groups in G02 (since this field did not include the updated photometry). $G^{3}C$v$_{11}$ does include the G23 field which was not publicly available in v$_{10}$. This field was also not used in \cite{driver2022a}. 

Furthermore, v$_{11}$ used the updated KiDS photometry with a brighter magnitude limit, decreasing from $r<19.8$ to $r<19.65$. Naturally this means that there are fewer galaxies and consequently fewer groups. This is what we see in Table \ref{tab:group_comp} across all multiplicity cuts; at $N\geq 3$ the number of groups across the equatorial fields drops from 8778 to 6980 ($\sim80\%$) of the groups found in v$_{10}$. This effect is also a function of multiplicity. Again, in the equatorial regions, for $N\geq2$, v$_{11}$ is $\sim85\%$ of v$_{10}$. But at $N\geq20$ this number is $\sim 38\%$. 

This can be explained by the combination of two effects which arise due to the change in magnitude limit: first, there is a cascading bias in play where if a larger group loses its members due to the brighter magnitude limit then it moves into a lower bin, replacing any group that might have itself cascaded down. Since there are fewer larger groups than smaller ones, smaller group bins are replenished preferentially; second, larger groups are more affected by the change in magnitude limit since they tend to have the faintest members, after the initial Brightest Cluster Galaxy (BCG) and core members have been identified. If we consider that there are also more faint galaxies than brighter ones, then the change in magnitude limit would very quickly remove members from large groups which have enough mass to contain so many faint galaxies, whereas smaller groups would be less affected.

A smaller effect would also contribute, where larger groups in the v10 catalogue could also be identified as multiple smaller groups in the v11 catalogue if they were erroneously identified as one large group because of faint interloper galaxies, which, having been removed from the sample, do not cause this bridging effect (see \citealt{lambert2020}). This would actually be a more accurately defined set of groups in this case, but we expect this effect to be relatively small. 

Regardless, due to the above reasons, it is worth noting that there are fewer large groups/clusters in v$_{11}$ than there are in v$_{10}$. Even including the G23 field, in this work, we still produce fewer groups at $N\geq5$.

\subsection{Comparison between Nessie SDSS groups and \citet{tempel2017}}
The comparison between the SDSS catalogue and \cite{tempel2017} is dependent purely on algorithmic choice since both group catalogues were built from the same galaxy catalogue. The differences between the two catalogues are smaller than between the two GAMA group catalogues, as can be seen in Table \ref{tab:sdss_comparison}. The scaling of the linking lengths between Nessie and \cite{tempel2017} differs fundamentally and therefore a perfect overlap is not expected.

The total number of groups across all redshifts and multiplicities differ by $\sim 4\%$. However, we notice that nearby ($z<0.02$) both catalogues agree at all multiplicities, but there is a difference as a function of redshift---\cite{tempel2017} begins to find lower multiplicity systems whereas Nessie finds higher ones.

This would suggest that the group finding algorithm used by \cite{tempel2017} preferentially identifies smaller multiplicity systems as opposed to Nessie which might prefer larger systems. Indeed, \cite{lambert2026} showed that the cost function that is used to tune the linking lengths for Nessie has a smaller penalty for combining groups than for splitting them up.

The different algorithmic choices between group finders is a topic that will be explored in a future work where we will compare various group finding algorithms using the same mock catalogue. However, for this work we simply state the differences for the reader and direct them to \cite{lambert2026} for more discussion between the two catalogues.

This SDSS comparison further supports the notion that the major differences between $G^{3}C$v$_{10}$ and $G^{3}C$v$_{11}$ are due to the photometric differences between the GAMA galaxy catalogues that were used (i.e., \citealt{liske2015} vs \citealt{driver2022}) as opposed to any algorithmic differences between this work and \cite{robotham2011}. Seeing that Nessie and \cite{tempel2017} can largely agree when using the same galaxy catalogue, despite fundamentally different approaches to the friends-of-friends algorithm, whilst at the same time having relatively major changes between $G^{3}C$v$_{10}$ and $G^{3}C$v$_{11}$, using a very slightly different algorithm, points to the input catalogue being the main source of the differences. 

\subsection{HMF}
Throughout all the HMF fits (Figs \ref{fig: hmf_gama}, \ref{fig: hmf_gama_sdss}, and \ref{fig: hmf_all}) the low-mass end of the fit deviates from the expected $\Lambda$CDM curve. This effect is worst in Fig. \ref{fig: hmf_gama}, which only included the GAMA catalogue and was also visible in \cite{driver2022a}. This early turnover is likely not physical and is instead being driven by incompleteness at the low-mass regime. A deeper survey would result in a more complete low-mass end. 

Although the low-mass end of the HMF fit to GAMA in Fig. \ref{fig: hmf_gama} does deviate, masses $\log_{10} (M/M_{\odot}) > 14$ follow the expected theoretical curve perfectly, essentially sitting on top of the line. We note, however, that when combined with the SDSS results in Fig. \ref{fig: hmf_gama_sdss} the high-mass end starts to deviate from \cite{Murray2021} slightly but does trace the curve reasonably well at lower masses, across two orders of magnitude, in the mass range $13<\log_{10}(M/M_{\odot})<15.5$. This turn-up at the high-mass end is driven by the overly high measurements in SDSS. These measurements themselves have large errors but still influence the fit.

The best fit, as was the case in \cite{driver2022a}, comes from combining GAMA, SDSS, and REFLEX II. These data points help constrain the high-mass end and we find that the curve closely follows the theoretical expectation from $\sim \log_{10}(M/M_{\odot}) > 13.3$. These results are very similar to those found by \cite{driver2022a} and, like that work, we also do not find any deviation from the expected $\Lambda$CDM model. However, in this work we again emphasize that we do not fit for $\phi^{*}$ and instead treat this as a nuisance parameter. Therefore, the strongest claim we can make is that the general shape of the HMF is as predicted in our unbiased mass range, and we make no claims regarding the actual number density that is expected. 

The final values for the MRP parameters in Table \ref{tab:mrp} show that GAMA and GAMA + SDSS were not enough to fully constrain the fit, as $M_*$ was set by the prior range rather than the data. When combined with the REFLEX II points, the parameters are constrained by the data. However, the measured values are offset from the predicted values of \cite{Murray2021} despite the curve fitting well. The deviation at the low-mass end will affect these measurements.

\section{Conclusion}
In this work we have updated the GAMA group catalogue using improved photometry from KiDS, and included the G23 field. This was done using Nessie \citep{lambert2026}, a generalized version of the original GAMA group finder. We find that there are fewer groups than in previous versions of the GAMA group catalogue, which is to be expected since the updated photometry resulted in a brighter magnitude limit. The differences between this catalogue and the previous catalogue are almost exclusively due to the change in the input galaxy catalogue due to the photometry as opposed to any algorithmic changes. 

In addition we also present a new SDSS catalogue, also using Nessie, built on the same galaxy input catalogue used in \cite{tempel2017}. We find that at low redshifts, these two catalogues agree in number density across all multiplicities, however, they start to diverge with increasing redshift with \cite{tempel2017} finding more smaller multiplicity systems whereas Nessie finds larger ones. Investigating the algorithmic differences which cause these differing results is to be discussed in a later paper.  

We combined both of these data sets with the HMF points of REFLEX II in \cite{Boringer2017}, to produce our best fit HMF attempt by following the method in \cite{driver2022a}, but instead choosing to treat $\phi^{*}$ as a nuisance parameter. The final best fit results were: $\log_{10}(M_*/M_{\odot}) = 12.60^{0.62}_{-0.79}$, $\alpha = -1.02_{0.24}^{0.27}$, $\beta = 0.37_{-0.06}^{0.07}$. Whilst these values deviate from the theoretical predictions of \cite{Murray2021}, the overall shape of the fitted curve closely follows the theoretical line for $\log_{10}(M/M_{\odot}) > 13.3$. 

From these results alone we cannot make any strong claims regarding the HMF, besides that the shape appears to follow the predicted $\Lambda$CDM curve. There are many different systematic errors that exist when constructing an empirical HMF. Besides the clear Eddington bias, which we have tried to address, there are more subtle systematics due to survey design, and, importantly, due to mass estimation; small, justifiable decisions for determining the mass can have a large systematic offset compared to other choices. Making an accurate enough measurement of the empirical HMF so that it can be used as a test of $\Lambda$CDM would require more care in addressing these systematics.

In addition, we have seen that wider and deeper surveys are needed to properly constrain the low-mass and high-mass ends. Fortunately, two such surveys are about to begin observations, namely the Wide Area VISTA Extragalactic Survey \citep[WAVES;][]{driver2019} and the 4MOST Hemisphere Survey \citep[4HS;][]{taylor2023a}. WAVES alone will already cover an area $\sim 5\times$ that of GAMA and about two magnitudes deeper. At the same time 4HS will cover the entire southern sky. The combination of these two surveys, as well as careful handling of the aforementioned systematics, might be enough to test $\Lambda$CDM in the future. Finally the DESI Bright Galaxy Survey \citep{Hahn2023} promises to observe nearly 10 million galaxies over 14 000 deg$^{2}$ which will also be invaluable in measuring the HMF.

\section*{Acknowledgements}

TSL, SD, and SB acknowledge support from the Australian Research Council (ARC) Laureate Fellowship scheme (FL220100191). ASGR acknowledges funding by the ARC Future Fellowship scheme (FT200100375).

Thanks to Elmo Tempel for discussions around mass estimates and providing us with his own module from his work.

GAMA is a joint European-Australasian project based around a spectroscopic campaign using the Anglo-Australian Telescope. The GAMA input catalogue is based on data taken from the Sloan Digital Sky Survey and the UKIRT Infrared Deep Sky Survey. Complementary imaging of the GAMA regions was obtained by a number of independent survey programmes including GALEX MIS, VST KiDS, VISTA VIKING, WISE, Herschel-ATLAS, GMRT and ASKAP providing UV to radio coverage. GAMA is funded by the STFC (UK), the ARC (Australia), the AAO, and the participating institutions. The GAMA website is https://www.gama-survey.org/ .
Based on observations made with ESO Telescopes at the La Silla Paranal Observatory under programme ID 179.A-2004.
Based on observations made with ESO Telescopes at the La Silla Paranal Observatory under programme ID 177.A-3016.

This work was supported by resources provided by the Pawsey Supercomputing Research Centre’s Setonix Supercomputer (https://doi.org/10.48569/18sb-8s43), with funding from the Australian Government and the Government of Western Australia.
\section*{Data Availability}

The Group catalogues built in this work will be available on the GAMA database (\url{https://www.gama-survey.org/dr4/schema/}) once they have passed the GAMA quality control process. The catalogues and any other data products are available via reasonable request to the TSL.



\bibliographystyle{mnras}
\bibliography{references} 




\appendix
\bsp	
\label{lastpage}
\end{document}